\documentclass[preprint, 11pt]{elsarticle}

\usepackage{url}
\usepackage{a4wide}

\usepackage{amsmath}
\usepackage{amsfonts}
\usepackage{amssymb}

\usepackage[mathscr]{eucal}

\usepackage{bm}

\usepackage{graphicx}

\usepackage{booktabs}

\setcitestyle{numbers, square, comma}

\newcommand{\R}{\mathbb{R}}

\newcommand{\bF}{\bm{F}}
\newcommand{\bU}{\bm{U}}

\usepackage{amsfonts}

\begin{document}

\begin{frontmatter}

\title{The spectral element method as an efficient tool for transient simulations of hydraulic systems}

\author[add1]{J.-F.~Mennemann\corref{cor1}}
\ead{mennemann@acin.tuwien.ac.at}

\author[add2]{L.~Marko}
\author[add1]{J.~Schmidt}
\author[add1]{W.~Kemmetm\"uller}
\author[add1,add2]{A.~Kugi}

\address[add1]{
Automation and Control Institute, 
Complex Dynamical Systems, 
TU Wien, 
Gusshausstrasse 27-29, 
1040 Vienna, 
Austria}

\address[add2]{ 
Christian Doppler Laboratory for Model-Based Process Control in the Steel Industry, 
TU Wien, 
Gusshausstrasse 27-29, 
1040 Vienna, 
Austria}

\cortext[cor1]{Corresponding author}

\begin{abstract}
This paper presents transient numerical simulations of hydraulic systems
in engineering applications using the spectral element method (SEM).
Along with a detailed description of the underlying numerical method, it is shown
that the SEM yields highly accurate numerical approximations at modest computational costs,
which is in particular useful for optimization-based control applications.
In order to enable fast explicit time stepping methods, the boundary conditions are imposed weakly
using a numerically stable upwind discretization.
The benefits of the SEM in the area of hydraulic system simulations are demonstrated in various examples
including several simulations of strong water hammer effects.
Due to its exceptional convergence characteristics, the SEM is particularly well suited to be used in
real-time capable control applications.
As an example, it is shown that the time evolution of pressure waves
in a large scale pumped-storage power plant can be well approximated using a
low-dimensional system representation utilizing a minimum number of dynamical states.
\end{abstract}


\end{frontmatter}

\section{Introduction}

Fast variations of the volume flow in a pipeline system give rise to pressure waves
propagating through the pressure lines and causing problems like noise, vibrations, or 
in the worst case, even pipe collapse.
These transient hydraulic effects constitute a major risk in various areas like cooling-water systems,
pumped-storage power plants or even in the feed lines of liquid rocket engines
\cite{wylie_1993, chaudhry_2014, dimatteo_2011, bandyopadhyay_2014}.

Wave propagation in pressure lines is well described by a system of hyperbolic partial differential equations
\cite{zielke_1968, wylie_1993, wu_1999, ghidaoui_2005, chaudhry_2014, guinot_2010}.
Various numerical methods for the simulation of hyperbolic systems are known in literature,
ranging from finite difference to finite volume
and spectral methods \cite{strikwerda_2004, leveque_2002, canuto_2006_fundamentals}.
Moreover, discontinuous Galerkin methods \cite{hesthaven_2007} represent an 
attractive alternative as they try to combine the advantages of the finite volume method with 
those of the finite element method.

However, the predominant method for the simulation of pressure waves in pipeline systems is the method
of characteristics (MOC) \cite{wylie_1993, wu_1999, adamkowski_2003, ghidaoui_2005, chaudhry_2014}.
It is based on the concept of the same name used to analyze and solve systems of hyperbolic partial
differential equations.
The MOC works best when all parameters characterizing the pressure lines are piecewise constant.
Under this condition, the MOC represents an extremely robust method which is able to propagate
shock waves with arbitrary steep gradients.
The main disadvantage results from the fact that the time step size, 
the spatial grid size and the local wave propagation speed are required to satisfy a (strict) 
Courant-Friedrichs-Lewy (CFL) condition.
To meet these requirements it is common practice to adjust the lengths or the local wave speeds 
of individual pipeline segments which may cause significant modeling errors.
Moreover, in case of non-constant coefficient functions a large number of spatial 
grid points is needed in order to compute satisfactory numerical approximations.

Besides the MOC, finite difference, finite volume, discontinuous Galerkin and spectral 
methods have been used to simulate transient hydraulic 
systems \cite{souza_1999, nicolet_2007, guinot_2000, leon_2007, koeppl_2013, chen_2013}.
In this context, it should be noted that numerous variations of spectral methods can be found in 
literature \cite{canuto_2006_fundamentals, canuto_2007_evolution}.
Ideally, spectral methods yield highly accurate solutions requiring only a minimum number of
degrees of freedom, which makes them attractive from a control applications point of view.
However, as spectral methods are formulated on a single interval or on a multidimensional cuboid, they
turn out to be too inflexible in many applications.

The spectral element method (SEM) takes this yet a step further and
applies a spectral method to every element of a properly partitioned computational domain.
In this manner, it is possible to combine the adaptivity of an ordinary finite element method
with the accuracy of a spectral method~\cite{patera_1984}.
In fact, the SEM could be interpreted as a special variation of a high-order finite element method.
However, a unique feature of the SEM is the quadrature rule, which relies on the same nodes as 
the definition of the element shape functions.
As a consequence, the mass matrix is a diagonal matrix, which enables the application of 
fast explicit time stepping methods.
Furthermore, thanks to the quadrature rule and the nodal interpolation property the assembling process
of nonlinear expressions can often be simplified significantly.

The SEM is widely used in elastic wave propagation and fluid dynamical 
problems~\cite{komatitsch_2005, bin_2015, beck_2014}.
In this paper, we apply the SEM to a system of hyperbolic partial differential equations 
frequently used to describe hydraulic systems in engineering applications.
As for all hyperbolic equations, 
special care needs to be taken with regard to the implementation of the boundary conditions.
Here, we choose to impose the boundary conditions weakly using a numerically stable upwind 
discretization.
As outlined above, the resulting mass matrix is a diagonal matrix.
Moreover, we show that the nonlinear pipe friction term can be easily taken into account.
Hence, the application of explicit time stepping strategies is very simple and yields extremely short 
computation times as needed in real-time control applications.

A minor drawback of the SEM might be seen in the fact that, in contrast to the MOC or 
discontinuous Galerkin methods, the SEM cannot propagate shock waves with arbitrary steep gradients.
Nonetheless, we will demonstrate that the SEM yields highly accurate solutions to rather strong water 
hammer problems using a modest number of spatial variables.
Additionally, it is shown that the underlying spatial semi-discretization can be integrated in time 
utilizing high-order explicit time-integration methods in combination with comparatively mild time 
step size restrictions.

Our aim is to demonstrate the benefits of the SEM in the area of 
transient hydraulic simulations, cf. also \cite{mennemann_2016}.
Moreover, we want to show that, due to its exceptional convergence properties, the SEM is 
particularly well suited to be used in optimal and model predictive control applications
utilizing direct transcription (full discretization) methods \cite{betts_2010, gerdts_2012}.

The paper is organized as follows:
In Section~\ref{sec:mathematical_modelling}, we introduce the underlying hyperbolic system which describes
wave propagation effects on a single pressure line.
Section~\ref{sec:numerical_method} covers the weak formulation, the spatial semi-discretization
and the implementation of Dirichlet boundary conditions.
In Section~\ref{sec:convergence_analysis}, we provide a careful numerical convergence analysis using 
several examples of increasing complexity. 
In this context, we also consider the above-mentioned strong water hammer problems.
Finally, Section \ref{sec:power_plant} is devoted to the simulation of wave effects in a 
pumped-storage power plant.
In particular, it is demonstrated that the SEM can be used to obtain low-dimensional 
semi-discretizations which are perfectly suited to be used in real-time capable control applications.

\section{Mathematical Modelling}
\label{sec:mathematical_modelling}

Wave propagation in pressure lines is well described in the literature
\cite{wylie_1993, wu_1999, ghidaoui_2005, chaudhry_2014, zielke_1968}.
Using mass and momentum balance equations it is possible to derive a system 
of hyperbolic partial differential equations.
A basic assumption is that the liquid flow is one-dimensional, i.e., the characteristic quantities 
are cross-section averaged~\cite{adamkowski_2003}.
In most applications, the average velocity of the fluid is small in comparison to the 
characteristic wave speeds $c$, which is why certain convective terms can be neglected.
Moreover, in many cases it is advantageous to consider the piezometric head $h$ instead 
of the pressure $p$.
Both quantities are related to each other in the form
\begin{equation}
\label{eq:relation_h_p}
h = \frac{p}{\rho g} + x_2,
\end{equation}
where $x_2 = x_2(z)$ denotes the height of the pressure line as function of the arclength $z$.
Furthermore, all simulations in this paper are based on the numerical values
$g=9.81\,\mathrm{m}/\mathrm{s}$ and $\rho = 1000\,\mathrm{kg}/\mathrm{m}^3$
for the gravitational acceleration and the density of water, respectively.

The time-evolution of the piezometric head $h$ and volume flow $q$ on a single pressure line
is well described by the hyperbolic system \cite{chaudhry_2014, wylie_1993}
\begin{equation}
\label{eq:system_hq}
\frac{\partial}{\partial t}
\begin{bmatrix}
h\\
q
\end{bmatrix}
+
\begin{bmatrix}
0 & c^2/(gA) \\
g A & 0
\end{bmatrix}
\frac{\partial}{\partial z}
\begin{bmatrix}
h\\
q
\end{bmatrix}
=
-
\begin{bmatrix}
0 \\
R q |q|
\end{bmatrix},
\end{equation}
where pipe friction is taken into account by the source term on the right hand 
side using a position-dependent function
\begin{equation*}
R = \frac{f_\lambda}{2 D A}.
\end{equation*}
Here, $f_\lambda$ denotes the Darcy-Weisbach friction factor and $D = 2 (A / \pi)^{1/2}$ 
is the diameter of the pressure line with $A$ the cross-sectional area.
Strictly speaking, this description of pipe friction is valid only for 
stationary pipe flow \cite{wu_1999}.
However, it is frequently used as a good approximation in transient simulations
as well \cite{wylie_1993, chaudhry_2014}.
A description of pipe frictional losses for non-stationary simulations is given 
in \cite{zielke_1968, wu_1999}.
In many situations of practical interest, the effect of pipe frictional losses is very small.
In particular, this will apply if fast transient simulations on a relatively short
time-scale are considered.
However, in stationary calculations, e.g., calculation of optimal stationary operating
points of pumped-storage power plants, 
even small contributions of pipe frictional losses 
might be important and thus need to be taken into account.

\section{Numerical Method}
\label{sec:numerical_method}

We now show how to apply the spectral element method to the hyperbolic system~\eqref{eq:system_hq}.
To this end, we first restrict ourselves to the case of a single pressure line described by the parameter
functions $A, c$ and $f_\lambda$ given on the computational domain $\Omega=[0,L]$.

\subsection{Weak formulation}

Introducing $\epsilon = gA/c^2$, $\mu= 1 / (gA)$ and $r = \mu R$,
system \eqref{eq:system_hq} can be written in conservative form (with source)
\begin{equation}
\label{eq:system_HQ_version_2}
\frac{\partial}{\partial t}
\begin{bmatrix}
\epsilon h\\
\mu q
\end{bmatrix}
+
\frac{\partial}{\partial z}
\bigg(
\begin{bmatrix}
0 & 1/\mu \\
1/\epsilon & 0
\end{bmatrix}
\begin{bmatrix}
\epsilon h\\
\mu q
\end{bmatrix}
\bigg)
=
-
\begin{bmatrix}
0 \\
r q |q|
\end{bmatrix},
\end{equation}
i.e.,
\begin{equation}
\label{eq:system_F}
\frac{\partial \bm{U}}{\partial t}
+ \frac{\partial \bm{F}(\bm{U})}{\partial z}
=
-\bm{f}(\bm{U}),
\end{equation}
where
\[
\bm{f}(\bm{U})
=
\begin{bmatrix}
0 \\
r \mu^{-1} U_2 |\mu^{-1} U_2|
\end{bmatrix}
\]
denotes the friction term as a function of the new variables
\[
\bm{U} =
\begin{bmatrix}
U_1\\
U_2
\end{bmatrix}
=
\begin{bmatrix}
\epsilon h\\
\mu q
\end{bmatrix}
\]
and the flux function is given by
\begin{equation}
\label{eq:flux_F}
\bF(\bU)
=
B \bm{U}
=
\begin{bmatrix}
0 & 1/\mu \\
1/\epsilon & 0
\end{bmatrix}
\begin{bmatrix}
U_1\\
U_2
\end{bmatrix}.
\end{equation}
Let $\mathcal{V}$ denote a suitable space of test functions on the interval $[0,L]$.
The weak formulation of \eqref{eq:system_F} reads
\begin{equation}
\label{eq:system_F_weak}
\int_0^L
\Big[
\frac{\partial \bm{U}}{\partial t} + \frac{\partial \bm{F}(\bm{U})}{\partial z}
+\bm{f} (\bm{U}) \Big] v \,dz 
= 
\bm{0} \;\;\;\textrm{for all}\;\;\; v \in \mathcal{V}.
\end{equation}
Using integration by parts, \eqref{eq:system_F_weak} becomes
\begin{equation}
\label{eq:system_F_weak_2}
\int_0^L \frac{\partial \bU}{\partial t} v \,dz
+
\bF(\bU) v \Big|_0^L
- \int_0^L \bF(\bU)  \frac{\partial v}{\partial z} \,dz
+ \int_0^L \bm{f} (\bU) v \,dz = \bm{0} \;\;\;\textrm{for all}\;\;\; v \in \mathcal{V}.
\end{equation}
Next, we replace the flux at the boundaries
by an appropriate numerical approximation.
In order to obtain a numerically stable and convergent scheme, we choose a so-called upwind flux.
In particular, we choose the Lax-Friedrichs flux \cite{canuto_2007_evolution} given as
\begin{equation*}
\bF^*(\bU^-, \bU^+)
=
\frac{1}{2} \big[ \bF(\bU^-) + \bF(\bU^+) \big]
- \frac{1}{2} |\lambda_\mathrm{max}(B)| \big( \bU^+ - \bU^- \big),
\end{equation*}
where
\[
\bU^-(z) = \lim_{\delta \rightarrow 0^+} \bU(z-\delta), \quad
\bU^+(z) = \lim_{\delta \rightarrow 0^+} \bU(z+\delta)
\]
and $|\lambda_\mathrm{max}(B)|$ denotes the maximum (in absolute value) eigenvalue of
the matrix $B$ defined in \eqref{eq:flux_F}.
Assuming that $\epsilon^- = \epsilon^+$ and $\mu^- = \mu^+$ holds true at the boundary points
$z=0$ and $z=L$, we find
\begin{equation}
\label{eq:numerical_flux}
\bF^*(\bU^-, \bU^+)
=
\frac{1}{2}
\begin{bmatrix}
(U_2^- + U_2^+) / \mu \\
(U_1^- + U_1^+) / \epsilon
\end{bmatrix}
-\frac{c}{2}
\begin{bmatrix}
U_1^+ - U_1^-\\
U_2^+ - U_2^-
\end{bmatrix}
=
\frac{1}{2}
\begin{bmatrix}
q^- + q^+\\
h^- + h^+
\end{bmatrix}
-
\frac{c}{2}
\begin{bmatrix}
\epsilon h^+ - \epsilon h^-\\
\mu q^+ - \mu q^-
\end{bmatrix}
\end{equation}
and hence, \eqref{eq:system_F_weak_2} yields
\begin{equation}
\label{eq:weak_formulation_final}
\begin{aligned}
\int_0^L \frac{\partial}{\partial t} \begin{bmatrix} \epsilon h\\ \mu q \end{bmatrix} v \,dz
= 
\int_0^L \begin{bmatrix} q \\ h \end{bmatrix} \frac{\partial v}{\partial z} \,dz
- \int_0^L 
\begin{bmatrix}
0 \\
r q |q|
\end{bmatrix} v \,dz
+
\begin{bmatrix}
F_{1}^*(0,t)\\
F_{2}^*(0,t)
\end{bmatrix} v(0)
-
\begin{bmatrix}
F_{1}^*(L,t)\\
F_{2}^*(L,t)
\end{bmatrix} v(L)
\end{aligned}
\end{equation}
for all $v \in \mathcal{V}$.
Here, $F_{1}^*(z,t)$ and $F_{2}^*(z,t)$ denote the components of the numerical 
flux $\bF^*$ evaluated at the position~$z$ and time $t$.

\subsection{Spatial discretization}

\begin{figure}
\begin{center}
\includegraphics[scale=1.0]{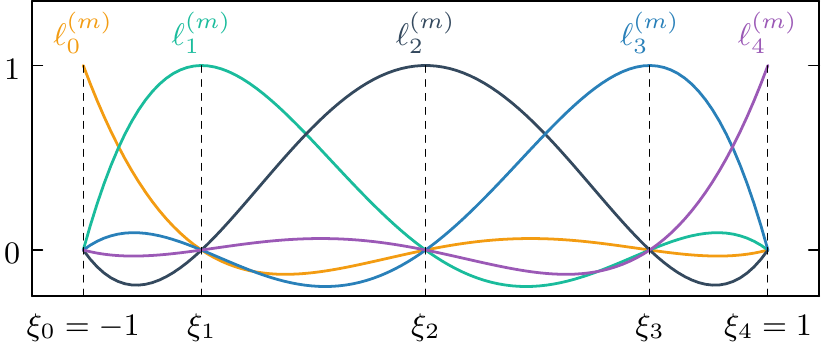}
\end{center}
\caption{
Element shape functions on the reference domain $\hat{\Omega}$ for $N_m = 4$.
}
\label{fig:shape_functions}
\end{figure}

In the following, we consider a partition of the computational domain into $M$ elements
\begin{equation*}
\Omega = \bigcup\limits_{m=1}^M Z_m
\end{equation*}
with $Z_m = [\bar{z}_{m-1}, \bar{z}_m]$ and $0 = \bar{z}_0 < \bar{z}_1 < \dots < \bar{z}_{M-1} < \bar{z}_M = L$.
Moreover, we define the finite-dimensional space
\[
\mathcal{V}_h 
= 
\{ 
v \in C^0(\Omega) : v_{|Z_m} \in \mathbb{P}_{N_m}, \,
m = 1, \dots, M 
\}.
\]
A precise description of a basis of $\mathcal{V}_h$ in the sense of the SEM can be 
found in the literature, see, e.g., \cite{canuto_2007_evolution}.
As in the case of the classical FEM the definition of the basis functions is given by means of
the element shape functions \cite{canuto_2007_evolution, larson_2013}, which
in turn are defined in terms of the Legendre-Gauss-Lobatto (LGL for short) nodes
$\xi_0, \dots, \xi_{N_m}$ on the reference domain $\hat{\Omega} = [-1,1]$.
The LGL-nodes are the zeros of
$
(1-\xi)^2 P_{N_m}'(\xi),
$
where $P_{N_m}'$ denotes the derivative of the Legendre polynomial $P_{N_m}$
of degree $N_m$, cf.~\cite{canuto_2006_fundamentals}.
Using the LGL nodes the element shape functions can be written as
\begin{equation}
\label{eq:element_shape_functions}
\ell_i^{(m)}(\xi) 
= 
\prod_{k=0, k\neq i}^{N_m}
\frac{(\xi-\xi_k)}{(\xi_i - \xi_k)},
\;\;\xi \in \hat{\Omega},\;\; i = 0, \dots, N_m
\end{equation}
with
\begin{equation}
\label{eq:interp_reference_element}
\ell_i^{(m)}(\xi_j) = \delta_{i,j},
\quad
j = 0, \dots, N_m.
\end{equation}
As an example, the element shape functions for $N_m = 4$ are shown in Fig.~\ref{fig:shape_functions}.

\begin{figure*}
\begin{center}
\includegraphics[scale=1.0]{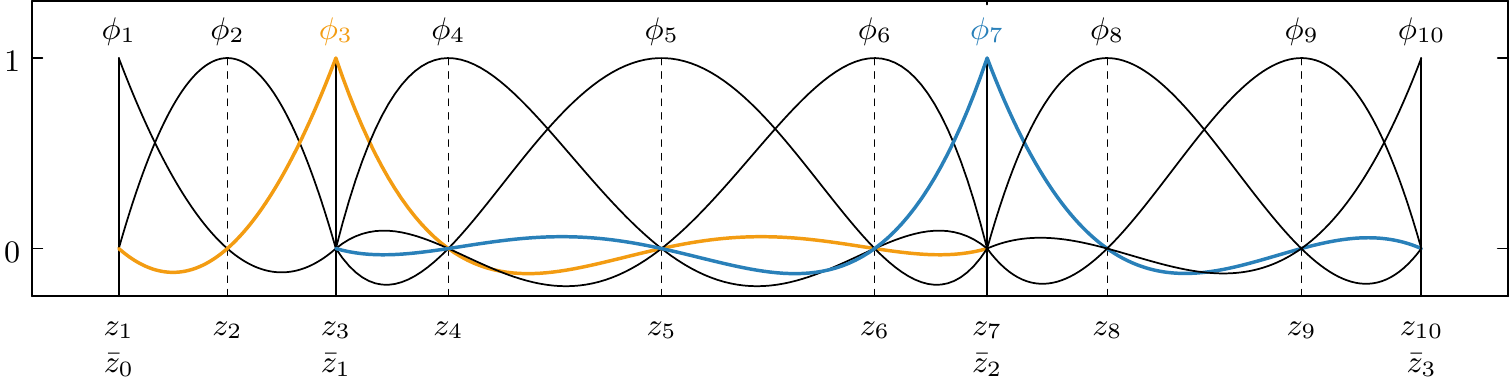}
\end{center}
\caption{
Global basis functions corresponding to a computational domain consisting of three elements.
According to the polynomial degrees $N_1 = 2$, $N_2 = 4$ and $N_3 = 3$ there are $10$ global
basis functions $\phi_1, \dots, \phi_{10}$ representing a basis of the ansatz space $\mathcal{V}_h$.
}
\label{fig:basis_functions}
\end{figure*}

By means of the map
\begin{equation}
\label{eq:element_transformation}
\Gamma_m(\xi) = \frac{\bar{z}_{m} - \bar{z}_{m-1}}{2} \xi + \frac{\bar{z}_m + \bar{z}_{m-1}}{2},
\;\; \xi \in \hat{\Omega}
\end{equation}
we may define the element basis functions
\[
\varphi_i^{(m)}(z) = \ell_i^{(m)}(\Gamma_m^{-1}(z)), \quad i=0,\dots,N_m
\]
on each element $Z_m = (\bar{z}_{m-1}, \bar{z}_m)$ with $m=1,\dots,M$.
The element basis functions are now used to construct the global basis functions
$\phi_1,\dots, \phi_J$.
To this end, all element basis functions are extended by zero outside of their domains.
Moreover, in order to ensure inter-element continuity at the nodes
$\bar{z}_1, \dots, \bar{z}_{M-1}$,
adjacent element basis functions are suitably combined
into one single basis function.
Fig.~\ref{fig:basis_functions} depicts an example
with three elements and the polynomial degrees $N_1 = 2$, $N_2 = 4$ and $N_3 = 3$.
In general, we construct
\begin{equation}
\label{eq:nr_of_basis_functions}
J = \sum_{m=1}^M N_m + 1
\end{equation}
global basis functions $\phi_j$, $j=1,\dots,J$ representing a basis for the ansatz space $\mathcal{V}_h$.

Using \eqref{eq:element_transformation}
the LGL nodes of the element shape functions can be
mapped to the domains of the corresponding elements.
Subsequently, we collect all nodes in
\begin{equation*}
\mathcal{Z} = \{z_1, \dots, z_J\},
\end{equation*}
where each inter-element node appears only once.
An important feature of the SEM is the nodal interpolation property:
\begin{equation}
\label{eq:interp_global}
\phi_i(z_j) = \delta_{i,j},
\quad
z_j \in \mathcal{Z},
\quad
i = 1, \dots, J.
\end{equation}

In the next step, we replace $h$ and $q$ in the weak formulation~\eqref{eq:weak_formulation_final} by the approximations
\begin{equation}
\label{eq:ansatz_hq}
h(z,t) \approx h_h(z,t) = \sum_{j=1}^J h_j(t) \phi_j(z)
\;\;\; \textrm{and} \;\;\;
q(z,t) \approx q_h(z,t) = \sum_{j=1}^J q_j(t) \phi_j(z).
\end{equation}
Moreover, we temporarily introduce
\begin{equation}
\label{eq:approximation_Q}
Q_h(z,t) = \sum_{j=1}^J Q_j(t) \phi_j(z), \quad Q_j(t) = q_j(t) |q_j(t)|, \quad j = 1, \dots, J
\end{equation}
to approximate the nonlinear expression $q(z,t) |q(z,t)|$ in the second component 
of~\eqref{eq:weak_formulation_final}.
In the sense of a Galerkin method, the space of test functions $\mathcal{V}$ is replaced by
the space of ansatz functions $\mathcal{V}_h$.
With the help of the mass matrix
\begin{equation*}
(M_\chi)_{i,j}
= 
\int_0^L \chi(z) \phi_i \phi_j \,dz,
\end{equation*}
the stiffness matrix
\begin{equation*}
S_{i,j} 
= 
\int_0^L \frac{\partial \phi_i}{\partial z} \phi_j \,dz,
\end{equation*}
as well as the vectors
\[
\bm{h} = \begin{bmatrix} h_1, \dots, h_J \end{bmatrix}^\top,
\quad
\bm{q} = \begin{bmatrix} q_1, \dots, q_J \end{bmatrix}^\top,
\quad
\bm{Q} = \begin{bmatrix} q_1 |q_1|, \dots, q_J |q_J| \end{bmatrix}^\top
\]
and
\[
\bm{e}_1 = \begin{bmatrix} 1, 0, \dots, 0 \end{bmatrix}^\top,
\quad
\bm{e}_J = \begin{bmatrix} 0, \dots, 0, 1 \end{bmatrix}^\top
\]
we obtain
\begin{subequations}
\label{eq:semi_discr_hq}
\begin{align}
M_\epsilon \frac{d \bm{h}}{dt}
&=
S \bm{q} + q_1^* \bm{e}_1 - q_J^* \bm{e}_J, \\
M_\mu \frac{d \bm{q}}{dt}
&=
S \bm{h} + h_1^* \bm{e}_1 - h_J^* \bm{e}_J - M_r \bm{Q},
\end{align}
where
\begin{equation}
\bm{Q} = \bm{q} |\bm{q}|
\end{equation}
\end{subequations}
and $\bm{q} |\bm{q}|$ denotes a pointwise multiplication of two column vectors.
In \eqref{eq:semi_discr_hq}, we have introduced the notations
\begin{equation*}
q_1^* := F_{1}^*(0,t), \quad
h_1^* := F_{2}^*(0,t), \quad
q_J^* := F_{1}^*(L,t), \quad
h_J^* := F_{2}^*(L,t),
\end{equation*}
which are motivated by the observation that $F_{1}^*(z,t)$ and $F_{2}^*(z,t)$ are approximations of
$q(z,t)$ and $h(z,t)$, respectively, cf.~\eqref{eq:flux_F}.
The values of $q_1^*, h_1^*, q_J^*$ and $h_J^*$ are used to realize the boundary conditions 
and will be specified in the subsequent sections.

Like in the case of classical finite element methods, the assembling process of the mass and stiffness matrices
is realized by means of the corresponding element mass and element stiffness matrices
\cite{kopriva_implementing_2009, larson_2013, pozrikidis_introduction_2014}.
In case of the SEM all integrals are approximated by the quadrature rule
\begin{subequations}
\label{eq:quadrature_rule}
\begin{equation}
\label{eq:quadrature_rule_sum}
\int_{-1}^{1} f(\xi) \,d\xi \approx \sum_{i=0}^{N_m} w_i f(\xi_i)
\end{equation}
using the weights \cite{canuto_2006_fundamentals}
\begin{equation}
\label{eq:quadrature_rule_weights}
w_i = \frac{2}{N_m (N_m+1)} \frac{1}{(P_{N_m}(\xi_i))^2}, \quad i = 0, \dots, N_m.
\end{equation}
\end{subequations}
Hence, the assembling process is based on the same nodes
$\xi_0, \dots, \xi_{N_m}$ as the definition of the element shape functions.
In fact, it can be easily verified that, as a result of the nodal interpolation property~\eqref{eq:interp_global}
and the quadrature rule \eqref{eq:quadrature_rule},
any mass matrix $M_\chi$ (irrespective of the coefficient function $\chi$) is a diagonal matrix \cite{canuto_2006_fundamentals}.
This property allows for the application of fast explicit time stepping methods to
the system of ordinary differential equations (ODEs) given in \eqref{eq:semi_discr_hq}.
The simplicity of the nonlinear pipe frictional loss term $M_r \bm{q} |\bm{q}|$
is even more remarkable.
Indeed, the mass matrix $M_r$ has to be computed only once.
This is in strong contrast to classical finite element methods which in the worst case scenario
would require to assemble a new mass matrix in each time step.
Finally, we would like to mention that the expression in~\eqref{eq:approximation_Q}
coincides with the nonlinear expression $q_h(z,t) |q_h(z,t)|$ for $z=z_j$, $j=1,\dots,J$.
Therefore,
$Q_h(z,t)$ and $q_h(z,t) |q_h(z,t)|$ are equivalent with respect to the
quadrature rule \eqref{eq:quadrature_rule}
and hence \eqref{eq:approximation_Q} does not introduce an additional disretization error.

\subsection{Dirichlet boundary conditions}

One of the most frequently used boundary conditions are Dirichlet boundary conditions
\begin{equation}
\label{eq:dirichlet_boundary_conditions}
h(0,t) = h_0(t) \textrm{ or } q(0,t) = q_0(t) \;\;\textrm{ and }\;\; h(L,t) = h_L(t) \textrm{ or } q(L,t) = q_L(t)
\end{equation}
for some given (potentially time dependent) functions $h_0(t), q_0(t), h_L(t)$ and $q_L(t)$.
In general, boundary conditions are realized using the values
$q_1^*, h_1^*$ and $q_J^*, h_J^*$ in \eqref{eq:semi_discr_hq} corresponding 
to the components of the numerical flux~\eqref{eq:numerical_flux}
\begin{equation}
\label{eq:numerical_flux_qh}
\begin{aligned}
q^* &= \frac{1}{2} (q^- + q^+) - \frac{g A}{2 c} (h^+ - h^-),\\
h^* &= \frac{1}{2} (h^- + h^+) - \frac{c}{2 g A} (q^+ - q^-)
\end{aligned}
\end{equation}
evaluated at the boundary points $z=0$ and $z=L$, respectively.

As an example let us consider the Dirichlet boundary condition $h(L,t) = h_L(t)$.
A simple strategy reported in literature of Discontinuous Galerkin methods
(cf. \cite{hesthaven_2007, busch_2011})
is given by
\begin{equation*}
q^+ = q^-
\textrm{ and }
h^+ = -h^- + 2 h_L(t),
\end{equation*}
i.e.,
\[
q_J^* = q^- - \frac{g A(L)}{c(L)} (h_L(t) - h^-) \;\; \mathrm{ and } \;\; h_J^* = h_L(t),
\]
where $q^-$ and $h^-$ have to be replaced by the last nodal values $q_J$ and $h_J$, respectively.
In the same way, all remaining boundary conditions of
\eqref{eq:dirichlet_boundary_conditions} can be realized, 
see Tab.~\ref{tab:dirichlet_boundary_conditions}.

\begin{table}
\centering
\begin{tabular}{lll}
\toprule
boundary condition & {numerical flux component $q^*$} 
& {numerical flux component $h^*$} \\
\midrule
{$h(0,t) = h_0(t)$}  & {$q_1^* = q_1 - \tfrac{g A(0)}{c(0)} (h_1 - h_0(t))$} & {$h_1^* = h_0(t)$} \\ 
{$q(0,t) = q_0(t)$}  & {$q_1^* = q_0(t)$}   & {$h_1^* = h_1 - \tfrac{c(0)}{g A(0)} (q_1 - q_0(t))$} \\
{$h(L,t) = h_L(t)$}  & {$q_J^* = q_J - \tfrac{g A(L)}{c(L)} (h_L(t) - h_J)$} & {$h_J^* = h_L(t)$} \\ 
{$q(L,t) = q_L(t)$}  & {$q_J^* = q_L(t)$}   & {$h_J^* = h_J - \tfrac{c(L)}{g A(L)} (q_L(t) - q_J)$}  \\
\bottomrule
\end{tabular}
\caption{
Numerical flux components $q^*$ and $h^*$ corresponding to different Dirichlet boundary conditions.
}
\label{tab:dirichlet_boundary_conditions}
\end{table}

\subsection{General remarks concerning the numerical simlations}

All simulations of this article are realized using the \texttt{Matlab} programming language.
In most examples, reference solutions are provided by the MOC.
In this context, it is important to note that due to the simple structure of the underlying 
algorithm, the MOC can be implemented very efficiently using \texttt{Matlab}-specific vectorization techniques.
However, as the MOC is a low-order numerical method, very small time step sizes $(\triangle t)_\mathrm{MOC}$
are required.
In its most basic form (without interpolations) the MOC requires the spatial step 
size $(\triangle z)_\mathrm{MOC}$ to satisfy the (strict) 
CFL condition $(\triangle z)_\mathrm{MOC} / (\triangle t)_\mathrm{MOC} = c$.
The situation becomes even worse
in pressure lines consisting of $K$ segments characterized by different wave speeds $c_k$, $k=1,\dots,K$
where the MOC needs to satisfy
\begin{equation}
\label{eq:CFL_MOC}
(\triangle z)_{\mathrm{MOC}}^{(k)} \big/ (\triangle t)_\mathrm{MOC} = c_k
\end{equation}
simultaneously for all segments $k=1,\dots,K$.
As a result, only certain combinations of spatial mesh sizes and time step sizes are possible.

A crucial step in the implementation of the SEM is the computation 
of the element mass and element stiffness matrices, which require the nodes
$\xi_0, \dots, \xi_{N_m}$ and the weights $w_0, \dots , w_{N_m}$ corresponding to the
quadrature rule given in \eqref{eq:quadrature_rule}.
Moreover, the first derivatives of the element shape functions $\ell_0^{(m)}, \dots, \ell_{N_m}^{(m)}$
need to be evaluated at $\xi_0, \dots, \xi_{N_m}$  in order to assemble the stiffness matrix.
Computing these numerical values with high precision is a nontrivial task, especially
if the polynomial degree $N_m$ is large.
We therefore make use of the \texttt{Matlab} routines given in~\cite{shen_spectral_2011},
which are numerically stable and efficient at the same time.

Using \eqref{eq:interp_global} and \eqref{eq:ansatz_hq} we immediately find
$h_h(z_j,t) = h_j$ and $q_h(z_j,t) = q_j$ 
for all $j=1,\dots,J$.
However, in order to compare solutions of the SEM with those of the MOC
we need to evaluate $h_h$ and $q_h$ at the (densely sampled) grid points associated to the MOC.
This interpolation process is realized using barycentric interpolation \cite{berrut_2004},
which is a fast and stable variant of Lagrange polynomial interpolation.
It is important to note that a direct evaluation of $h_h$ and $q_h$ based on
formula~\eqref{eq:element_shape_functions} is numerically unstable and would ruin all numerical 
convergence rates reported below.

Finally, we would like to mention that in the graphical representations
we always show the spatial distributions of the pressure $p$ instead of the piezometric head $h$.
However, since the numerical implementation of the SEM and the MOC are based on 
\eqref{eq:system_hq} which is formulated in terms of the piezometric head $h$ and the
volume flow $q$, we show numerical errors and relative differences in terms of the piezometric head $h$.
The numerical differences in terms of $q$ would look very similar and yield the
same convergence rates.

\section{Numerical convergence analysis}
\label{sec:convergence_analysis}

In the following, we consider several numerical examples of increasing complexity.
The simulations in this section are limited to a single horizontal pressure line.
Without loss of generality, we set $x_2(z) = 0$ for all $z \in \Omega$, $\Omega = [0,L]$.
The length of the pressure line $L$  and its diameter $D$ are similar to those
used in \cite{chen_2013}.

\subsection{A mathematical benchmark problem}
\label{sec:benchmark_problem}

In the first example, we consider constant coefficient functions $D,c>0$ and $f_\lambda=0$.
In particular, we are interested in computing the time evolution corresponding to the initial values
\begin{align*}
h(z,t=0) = h_s(z), \quad q(z,t=0) = q_s(z)
\end{align*}
for $z \in \R$ and $t > 0$.
Using the characteristic decomposition
\begin{align*}
\partial_t w_1 + c \partial_z w_1 = 0, \quad w_1 &= \frac{1}{2} \Big[ q + \frac{g A}{c} h \Big], \\
\partial_t w_2 - c \partial_z w_2 = 0, \quad w_2 &= \frac{1}{2} \Big[ q - \frac{g A}{c} h \Big],
\end{align*}
it follows immediately that
\begin{equation}
\label{eq:reference_solution}
h(z,t) 
= 
\frac{c}{g A}
\big[ w_{1,s}(z-ct) - w_{2,s}(z+ct) \big]
\;\;\; \textrm{and} \;\;\;
q(z,t)
=
w_{1,s}(z-ct) + w_{2,s}(z+ct)
\end{equation}
solve \eqref{eq:system_hq} for all $z,t \in \R$.
Here, $w_{1,s}$ and $w_{2,s}$ denote the initial values given by 
$w_{1,s}(z) = [q_s(z) + (g A / c) h_s(z)] / 2$
and  $w_{2,s}(z) = [q_s(z) - (g A / c) h_s(z)] / 2$, respectively.

At the boundaries of $\Omega$ we implement the following pair of transparent boundary conditions:
\begin{equation}
\label{eq:transparent_boundary_conditions}
q(0,t) = -\frac{g A}{c} h(0,t) \;\;\textrm{ and } \;\; q(L,t) = \frac{g A}{c} h(L,t).
\end{equation}
Obviously, these boundary conditions are of different nature compared to those 
listed in Tab.~\ref{tab:dirichlet_boundary_conditions}.
However, we may define the functions $q_0(t) = - (g A / c) h_1$ and $q_L(t) = (g A / c) h_J$
with the first and last nodal value $h_1$ and $h_J$, respectively.
Subsequent application of the second and fourth line of Tab.~\ref{tab:dirichlet_boundary_conditions}
yields
\begin{equation}
\label{eq:silver_mueller}
q_1^* = -\frac{g A}{c} h_1, \; h_1^*= -\frac{c}{gA} q_1 \;\; 
\mathrm{ and } \;\; 
q_J^* = \frac{g A}{c} h_J, \; h_J^*=\frac{c}{gA} q_J.
\end{equation}
We note that these equations coincide exactly with the Silver-M\"uller boundary conditions
(reduced to one spatial dimension) presented in \cite{busch_2011}.
Below we will see that \eqref{eq:silver_mueller} can be used
in combination with explicit time stepping methods and reasonable time step sizes~$\triangle t$.
However, we found that a simple modification of \eqref{eq:silver_mueller}
results in a stable discretization for even larger time step sizes.
To this end, we require \eqref{eq:transparent_boundary_conditions} to be satisfied
for the numerical flux values $q^*$ and $h^*$, i.e. we define
$q_0(t) = -(g A / c) h_1^*$ and $q_L(t) = (g A / c) h_J^*$.
Application of the second and fourth line in Tab.~\ref{tab:dirichlet_boundary_conditions}
yields
\begin{equation}
\label{eq:transparent_boundary_conditions_discr}
q_1^* = \frac{1}{2} \Big( q_1 - \frac{g A}{c} h_1 \Big), \,
h_1^* = \frac{1}{2} \Big( h_1 - \frac{c}{gA}  q_1 \Big)
\; \mathrm{ and } \; 
q_J^* = \frac{1}{2} \Big( q_J + \frac{g A}{c} h_J \Big), \, 
h_J^* = \frac{1}{2} \Big( h_J + \frac{c}{g A} q_J \Big).
\end{equation}

\begin{figure}
\begin{center}
\includegraphics[scale=0.925]{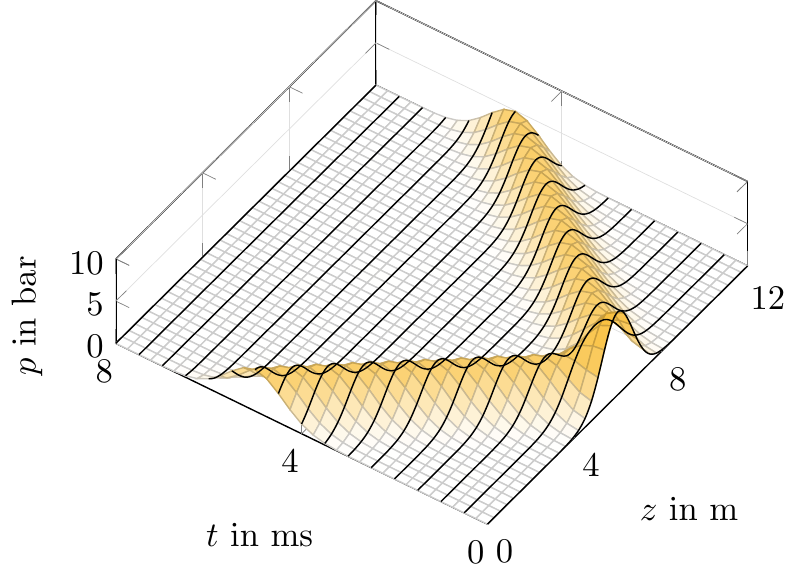}
\includegraphics[scale=0.925]{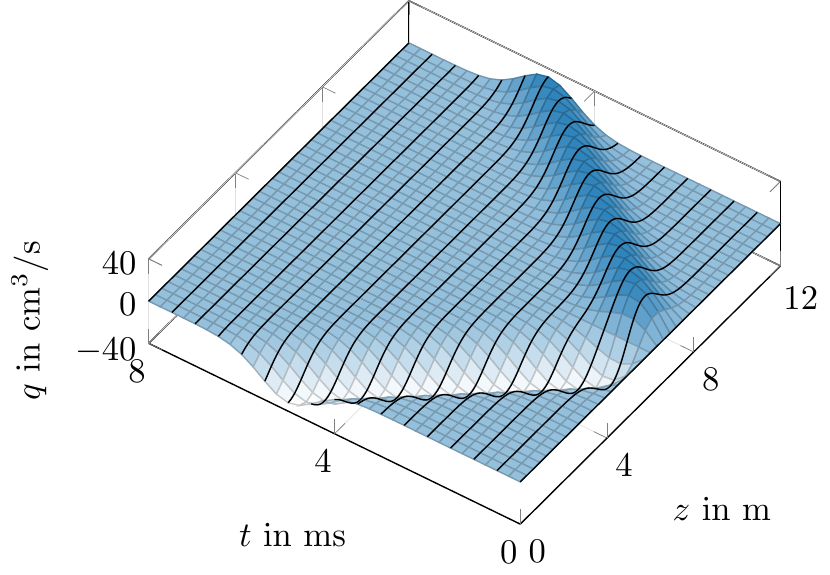}
\end{center}
\caption{
Time evolution of $p$ and $q$ corresponding to the mathematical benchmark problem
considered in Sec.~\ref{sec:benchmark_problem}.
}
\label{fig:transparent_time_evolution}
\end{figure}

For the numerical experiment we choose $D=0.01$m, $c=1200\,\mathrm{m}/\mathrm{s}$ and
$L=12$\,m.
Furthermore, we consider the initial value distributions
\[
h_s(z) = \alpha \exp(- \beta (z-L/2)^2), \quad q_s(z) = 0
\]
for $z \in \Omega$ with $\alpha = 100$\,m and  $\beta = 1$.

In the first numerical simulation, the computational domain is partitioned into only $M=10$ elements 
of equal size and a polynomial degree of $N_m=5$ is used for all elements $m=1,\dots,M$.
In order to propagate the numerical solution in time
we employ the classical Runge-Kutta method of order $4$ \cite{butcher_2016}.
An illustration of the time evolution of $p$ and $q$ 
for $t \in [0,T]$ with $T=8$\,ms
is given in Fig.~\ref{fig:transparent_time_evolution}.
Using the numerical flux components given in \eqref{eq:transparent_boundary_conditions_discr} we
may utilize a time step size of $\triangle t = 0.2$\,ms which corresponds to only 
$40$ time steps for the whole simulation.
In contrast, a time step size of $\triangle t = 0.05$\,ms is needed to ensure
numerical stability if the numerical flux values are chosen according to
\eqref{eq:silver_mueller}.

\begin{figure}
\begin{centering}
\includegraphics[width=15cm]{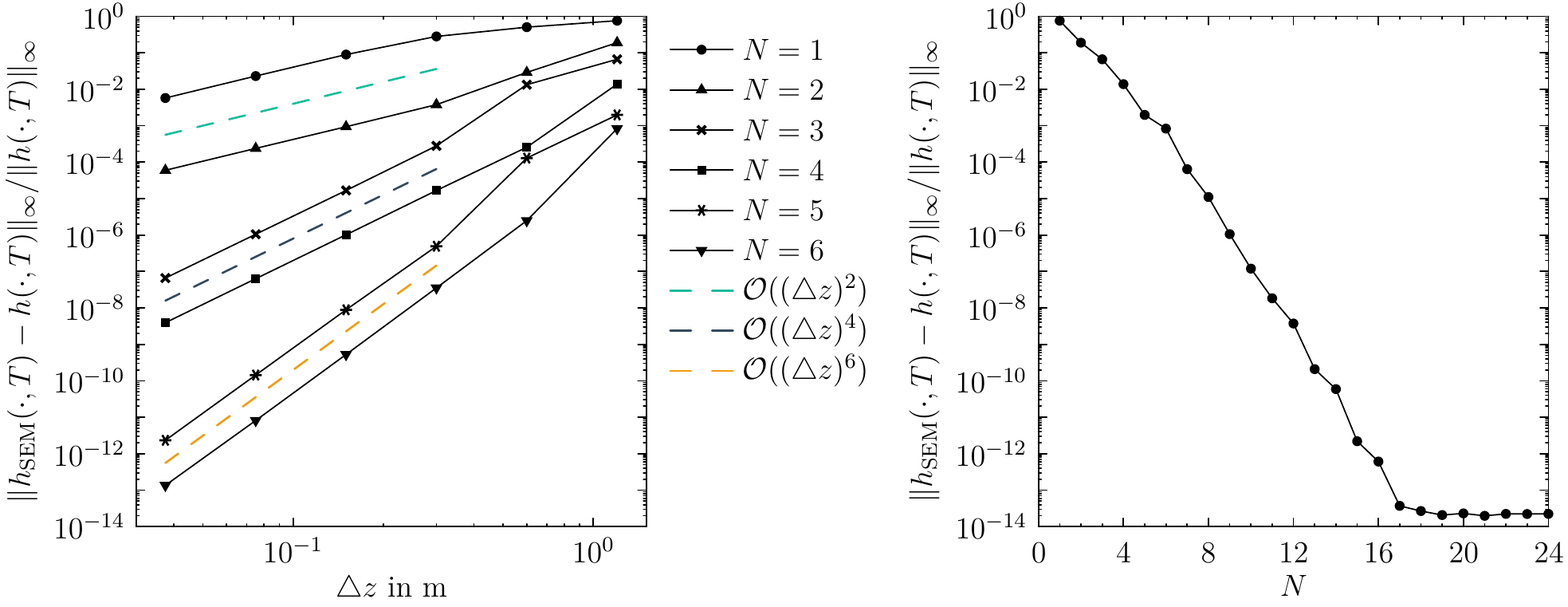}
\caption{
Relative errors corresponding to the mathematical benchmark problem evaluated at $T=5$\,ms.
Left column:
During a series of simulations, the polynomial degrees $N_m = N$  are kept constant for all elements $m=1, \dots, M$
while the computational domain is divided into decreasing element sizes $\triangle z$.
The convergence rates are of order $N+1$ for odd degree and $N$ for even degree.
Right column: The computational domain is partitioned into $M=10$ elements of equal size.
In each simulation, the polynomial degree $N_m = N$ is successively increased for all 
elements $m=1,\dots,M$.
The numerical error decreases at a spectral, i.e., exponential convergence rate.
}
\label{fig:transparent_convergence_sem}
\end{centering}
\end{figure}

As a next step we want to compute numerical convergence rates with respect to the 
size of the elements.
To this end, we divide the computational domain into $M$ elements of the same size $\triangle z$.
Moreover, we employ the same polynomial degree $N_m = N$ for all $m=1, \dots, M$.
During a series of simulations we keep the polynomial degree constant whereas the computational domain
is divided into decreasing element sizes $\triangle z$.
In order to keep the time discretization error as low as possible
we use the \texttt{Matlab} function 
\texttt{ode45}~($\texttt{RelTol} = 2.5\times 10^{-14}$). 
In this specific example, an exact reference solution is available and can 
be computed using \eqref{eq:reference_solution}.
The relative errors are evaluated at time $T=5$\,ms, i.e., exactly when half of the wave packet
has already left the computational domain.
In this manner, it is guaranteed that the effect of the boundary conditions is fully
included in the numerical convergence rates depicted 
in the left column of Fig.~\ref{fig:transparent_convergence_sem}.
It can clearly be seen that the convergence rates are of order $N+1$ for odd degree
and $N$ for even degree.
In this context, it is worth noting that such odd-even phenomena have already been reported in
literature of discontinuous Galerkin methods (DGM) \cite{debasabe_2008, riviere_2008}
and that the way we implement the boundary conditions is based on the same principles
as in the DGM.

The effect reported above might be interesting from a theoretical point of view, however, 
we will now demonstrate that for realistic element sizes $\triangle z$
it makes hardly any difference whether one uses odd or even polynomial degrees.
For this purpose, we partition the computational domain into $M=10$ elements of the same size.
In each simulation, the polynomial degree $N_m = N$ is successively increased for all 
elements $m=1,\dots,M$
and the relative errors are gathered in the right column of Fig.~\ref{fig:transparent_convergence_sem}.
The resulting graph perfectly agrees with the anticipated spectral, i.e., exponential convergence rate.
Furthermore, there is a significant improvement of the accuracy irrespective of whether 
one increases $N$ starting from an even or an odd polynomial degree.

\subsection{Water hammer simulation 1}
\label{sec:water_hammer_simulation_1}

We consider the same pressure line as in the previous example.
However, the boundary condition at the left boundary point $z=0$ is replaced with a nonhomogenous 
Dirichlet boundary condition $h(0,t) = h_0$, where $h_0 = p_0 / (\rho g)$ for some given pressure
$p_0$.
At the same time the boundary condition corresponding to the right boundary point $z=L$ is replaced
with the boundary condition
\begin{equation*}
q(L,t) = \alpha_v A_v \sqrt{ 2 / \rho } \operatorname{sign}(p(L,t)-p_v)
\,
\sqrt{\left| p(L,t)- p_v \right| } \, u(t),
\end{equation*}
which describes the turbulent volume flow through a valve.
Here, $p_v$ denotes the pressure at the valve outlet,
$A_v$ is the cross-sectional area,
$\alpha_v$ is referred to as the contraction coefficient,
and $u(t)$ describes the control input  related to the opening of the valve. 
Assuming that $p(L,t) > p_v$ holds true for all times $t$, we may write
\begin{equation}
\label{eq:boundary_conditions_valve}
q(L,t) = C_v \sqrt{ h(L,t) - h_v } u(t)
\end{equation}
with $C_v = \alpha_v \sqrt{2 g} A_v$ and $h_v = p_v / (\rho g)$.

In a water hammer simulation, the initial values of $h$ and $q$
are given by the stationary solution of \eqref{eq:system_hq} and the
control input $u$ is decreased from a positive value $u_0$ 
at time $t=0$ to $u_{T_c}=0$ at time $t=T_c$ in a strictly monotonous manner.
In order to prevent a distortion of the convergence rates reported below,
the control input needs to be a sufficiently smooth function.
Therefore, we employ
\begin{equation}
\label{eq:raised_cosine_high_order}
u(t) =
\begin{cases}
1, &\mbox{if } t < 0, \\
\sigma(t \pi/T_c) &\mbox{if } t \in [0, T_c], \\
0, &\mbox{if } t > T_c,
\end{cases}
\end{equation}
where $\sigma$ denotes the eighth order sharpened raised cosine filter \cite{canuto_2006_fundamentals}
\[
\sigma(\theta) = \sigma_0^4(\theta) [35 - 84 \sigma_0(\theta) + 70 \sigma_0^2(\theta) - 20 \sigma_0^3(\theta)]
\]
with
\[
\sigma_0(\theta) = (1 + \cos(\theta)) / 2, \quad \theta \in [0, \pi].
\]
We note that $u$ is seven times continuously differentiable in $\R$.

\begin{figure}[htb]
\begin{centering}
\includegraphics[scale=1]{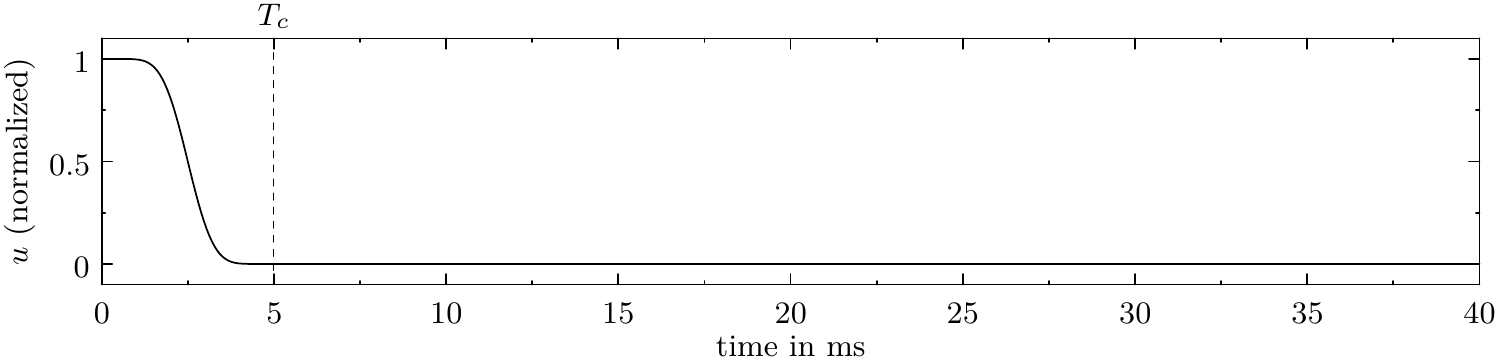}
\caption{
Control input $u$ of the valve 
corresponding to the first and second water hammer simulation.
}
\label{fig:raised_cosine_8th_order}
\end{centering}
\end{figure}

The first boundary condition at the left boundary point $z=0$ can be implemented easily using
the first line of Tab.~\ref{tab:dirichlet_boundary_conditions}.
Similar to the previous example we consider two methods to realize the second boundary 
condition in \eqref{eq:boundary_conditions_valve}.
To this end, we consider the components of the numerical flux
\begin{equation}
q_J^* = q_L(t), \quad h_J^* = h_J - \frac{c}{gA} (q_L(t) - q_J)
\label{eq:q_L_star_h_L_star}
\end{equation}
corresponding to the last line of Tab.~\ref{tab:dirichlet_boundary_conditions}.
In case of the first method, \eqref{eq:boundary_conditions_valve}
is evaluated using the last nodal value of $h$:
\begin{equation}
q_L(t) = C_v \sqrt{h_J - h_v} u.
\label{eq:q_valve_simple}
\end{equation}
In case of the second approach, we require \eqref{eq:boundary_conditions_valve}
to be satisfied for the numerical flux values $q_J^*$ and $h_J^*$ itself. 
Therefore, $q_L(t) = q_J^*$ is given implicitly via
\begin{equation}
q_J^* = C_v \sqrt{h_J^* - h_v} u.
\label{eq:q_valve_advanced}
\end{equation}
In a numerical implementation, $h_J^*$ is replaced according to the second expression in 
\eqref{eq:q_L_star_h_L_star} and hence we obtain a quadratic equation in $q_J^*$.
However, as we will see below, 
the numerical effort caused by solving an additional quadratic equation is negligibly
small compared to the improvements regarding the stability of the resulting numerical scheme.

The following numerical experiments are based on the parameter values
$p_0 = 120$\,bar, $p_v = 100$\,bar, $\alpha_v = 0.7$ and $A_v = A_L/5$,
where $A_L$ denotes the cross-sectional area at $z=L$.
The control input is given according to~\eqref{eq:raised_cosine_high_order} 
using a valve closure time of $T_c=5$\,ms, see Fig.~\ref{fig:raised_cosine_8th_order}.
All other parameters $D=0.01$m, $c=1200\,\mathrm{m}/\mathrm{s}$, $L=12$\,m and 
$f_\lambda=0$ remain unchanged.

\begin{figure}
\begin{center}
\includegraphics[scale=0.75]{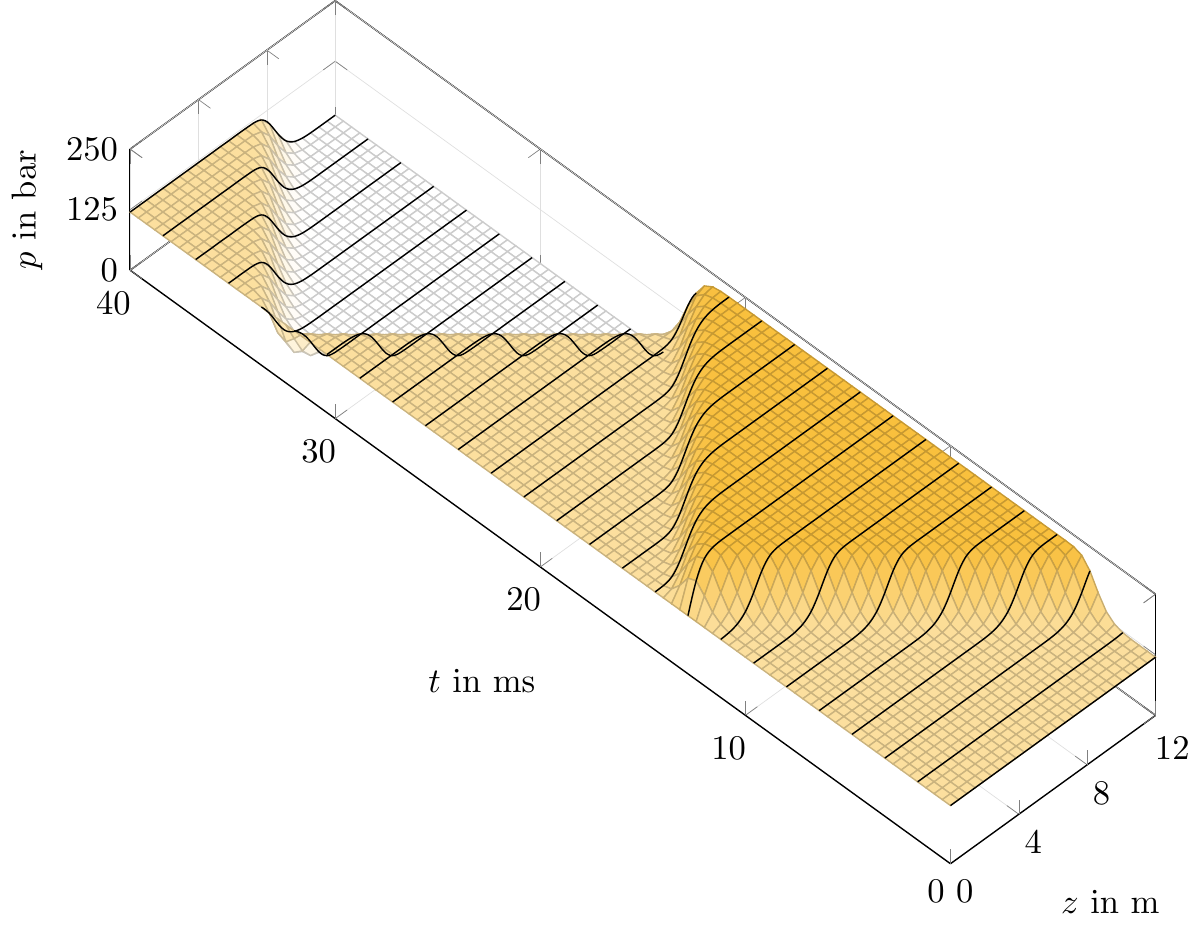}
\hspace{-3.5cm}
\includegraphics[scale=0.75]{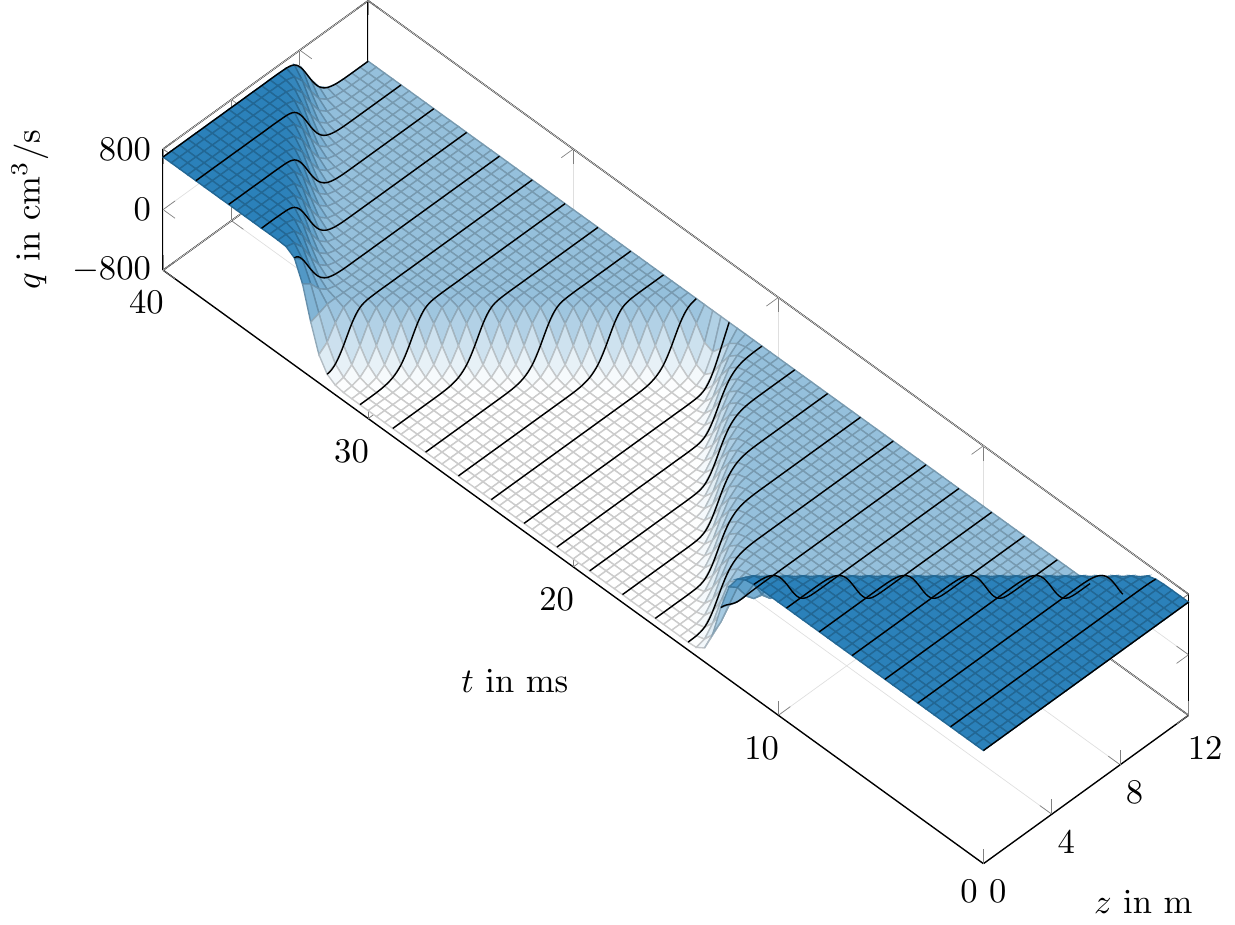}
\end{center}
\caption{
Time evolution of $p$ and $q$ corresponding to the first water hammer problem.
}
\label{fig:water_hammer_1_time_evolution}
\end{figure}

In the first numerical experiment, we partition the computational domain into $M=10$ elements
of equal size and a polynomial degree of $N_m = 5$ is employed on each element $m=1\dots,10$.
For the implementation of the valve boundary condition, we employ the numerical flux values according to
the implicit formulation~\eqref{eq:q_valve_advanced}.
The time integration is realized using the classical Runge Kutta method in combination with a 
time step size of $\triangle t = 0.2$\,ms which corresponds to only $200$ time steps for the whole 
time evolution shown in Fig.~\ref{fig:water_hammer_1_time_evolution}.
In contrast, if the valve boundary condition is realized using~\eqref{eq:q_valve_simple},
a time step size of $\triangle t = 0.025$\,ms is needed to ensure numerical stability.
Thus, a minor modification of the discretization scheme improves the computational efficiency
by a factor of almost ten.

\begin{figure}[htb]
\begin{center}
\includegraphics[scale=0.9]{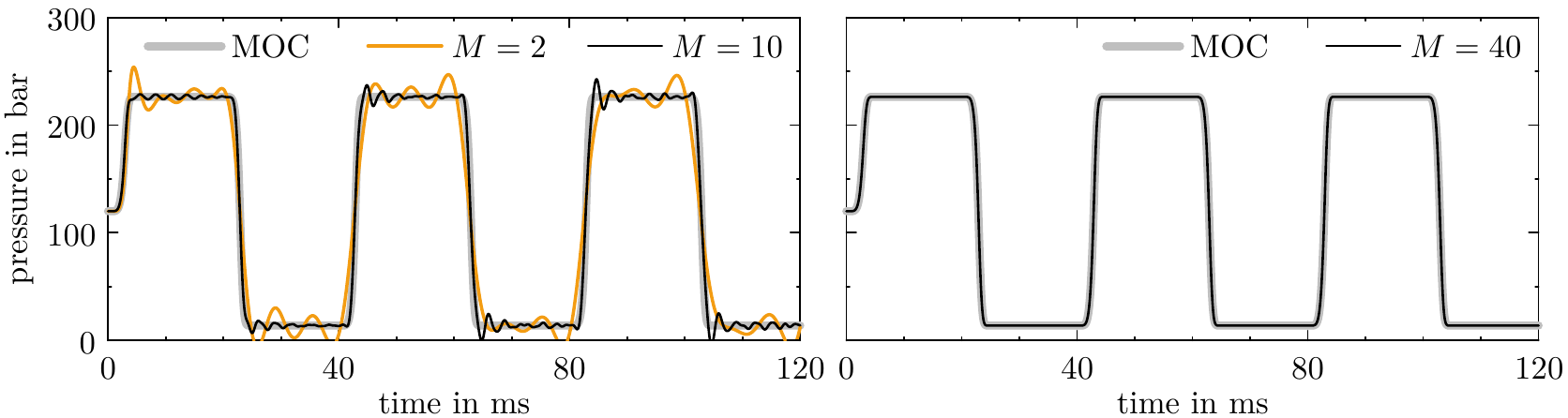}
\end{center}
\caption{
Time evolution of the pressure at $z=L$ corresponding to the first water hammer problem
using different numbers of elements $M$
with fixed polynomial degrees $N_m = 3$, $m=1, \dots,M$.
}
\label{fig:water_hammer_1_signatures_of_M}
\end{figure}

Next, we take a look at the water hammer signature, i.e., the time evolution of the pressure at $z=L$.
In particular, we compare the water hammer signatures obtained by means of the SEM with those
obtained using the MOC.
In order to satisfy the CFL condition of the MOC (for all considered element sizes), a
comparatively small time step size of $\triangle t = 0.025$\,ms is used for the time integration.
Note that this restriction only applies to the MOC. 
In fact it would be possible to use a much larger time step size for the time integration of the SEM.
As before, the time integration of the SEM is based on the classical Runge-Kutta method.

In Fig.~\ref{fig:water_hammer_1_signatures_of_M}, we show the results of the SEM
using a different number of equally sized elements $M$ with fixed 
polynomial degrees $N_m=3$ for all $m=1,\dots,M$.
While for small $M$ we observe significant overshoots similar to the Gibbs phenomenon,
these overshoots decrease rapidly with increasing $M$.
In fact, for $M=40$ we perfectly recover the water hammer signature computed by the MOC.

\begin{figure}[htb]
\begin{center}
\includegraphics[scale=0.9]{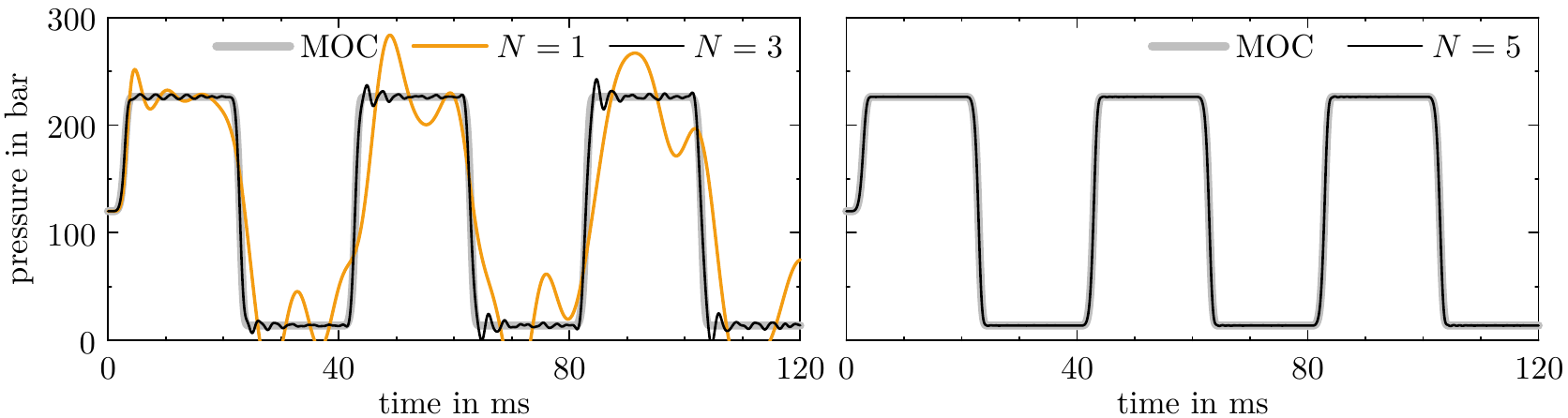}
\end{center}
\caption{
Time evolution of the pressure at $z=L$ corresponding to the first water hammer problem
using a fixed number of elements $M=10$ with different polynomial degrees 
$N_m = N$, $m=1, \dots,M$.
}
\label{fig:water_hammer_1_signatures_of_N}
\end{figure}

The results of a complementary situation are depicted in Fig.~\ref{fig:water_hammer_1_signatures_of_N}.
Here, we consider a fixed number of elements $M=10$ but different polynomial degrees 
$N_m = N$ with $m=1, \dots,M$.
It can clearly be seen that, as the polynomial degree is increased, the solutions corresponding to the
SEM converge quickly towards the solution obtained by the MOC.

Considering very similar simulations in \cite{chen_2013},
the results presented above are quite remarkable.
Using only $M=10$ elements in combination with a polynomial degree of $N=5$, we are able 
to perfectly recover the water hammer signature computed by the MOC.
This is in strong contrast to the Chebyshev super spectral viscosity method \cite{chen_2013}
which, by means of an additional viscosity operator,
is able to counteract the Gibbs-like phenomenon. 
However, at the same time, the viscosity operator leads to
a significant smoothing of the original water hammer signature.
In this context, it is important to note that the system sizes of both discretizations are 
comparable.
Moreover, the water hammer effect considered here is very similar in terms
of the ratio of the valve reflection time $(2 L / c)$ to the valve closure time $T_c$.

Finally, we compute numerical convergence rates corresponding to the SEM
as a function of the element size $\triangle z$ and the polynomial degree $N_m$.
The time evolution is again based on the \texttt{Matlab} function 
\texttt{ode45}~($\texttt{RelTol} = 2.5\times 10^{-14}$)
which allows to keep the time discretization error as low as possible.

Due to the nonlinear valve boundary condition~\eqref{eq:boundary_conditions_valve},
an exact reference solution would be extremely difficult to obtain.
We therefore compute a numerical reference solution using the MOC.
The relative differences will be evaluated at time $T = 20$\,ms.
Since the MOC is only a low order method, an extremely small time step size of 
$(\triangle t)_\mathrm{MOC} = 0.00025$\,ms is needed in order to obtain sufficiently 
accurate solutions, and as a result of the CFL condition~\eqref{eq:CFL_MOC}, 
the number of spatial grid points $M_\mathrm{MOC} = 40\,000$ is quite large.
Thus, computing a reference solution at time $T=20$\,ms is a rather
time-consuming process.

\begin{figure}
\begin{centering}
\includegraphics[width=155mm]{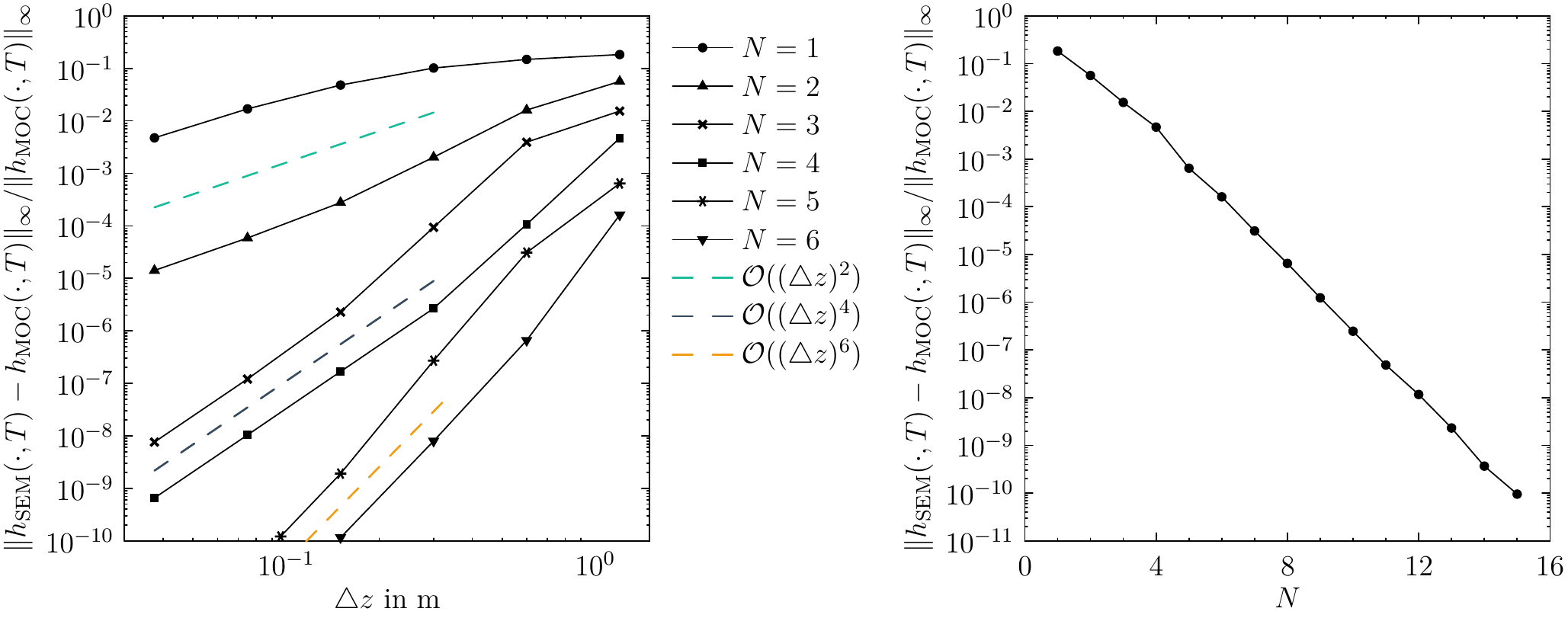}
\caption{
Relative differences corresponding to the first water hammer problem 
evaluated at $T=20$\,ms.
Left column:
In a series of simulations, 
the computational domain is divided into decreasing element sizes $\triangle z$
while the polynomial degrees $N_m = N$  are kept constant for all elements $m=1, \dots, M$.
Right column: The computational domain is partitioned into $M=10$ equally sized elements.
In each simulation, the polynomial degree $N_m = N$ is successively increased for all 
elements $m=1,\dots,M$.
It can be clearly seen that the numerical error decreases at an exponential convergence rate.
}
\label{fig:convergence_sem}
\end{centering}
\end{figure}

The left column of Fig.~\ref{fig:convergence_sem} shows the relative differences as a function of the
element size $\triangle z$ for fixed polynomial degrees $N_m = N$ for all $m=1,\dots,M$.
As we can see, we obtain the same numerical convergence rates as in the mathematical benchmark 
problem (cf. Fig.~\ref{fig:transparent_convergence_sem}) exhibiting the same odd-even phenomenon.
However, we will again show that in realistic scenarios, where the number of elements is much smaller,
this odd-even phenomenon is not an issue.
To this end, let us consider a subdivision of the computational domain into $M=10$ equally sized elements.
While the number of elements is kept constant we compute the relative differences for
increasing polynomial degrees $N_m=N$ with $m=1, \dots, M$.
The results are gathered in the right column of Fig.~\ref{fig:convergence_sem}.
Like in the mathematical benchmark problem we observe a nearly perfect spectral
convergence rate.
Moreover, no staircase effect is visible and, therefore, the improvement of
accuracy is not dependent on whether we increase $N$ starting from an odd or an even 
polynomial degree.

\subsection{Water hammer simulation 2}

\begin{figure*}
\begin{center}
\includegraphics[scale=1.0]{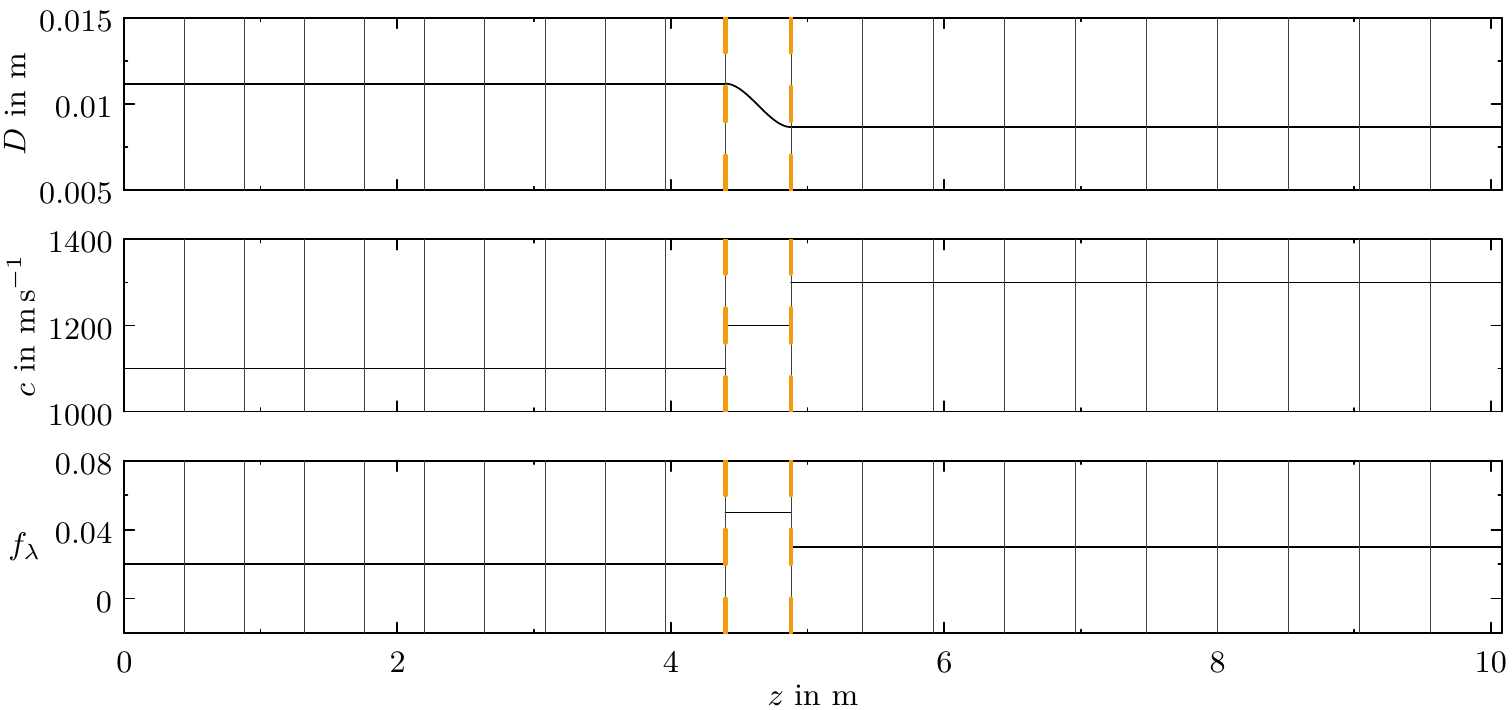}
\end{center}
\caption{
Parameters of the pressure line used in the second water hammer problem.
The pressure line consists of three segments (separated by dashed orange lines)
as needed to resolve the abrupt changes in the coefficient functions.
In the given example, the segments are subdivided into $M_1=10$, $M_2=1$ and $M_3=10$ elements.
}
\label{fig:figure_water_hammer_2_parameters}
\end{figure*}

So far, we neglected the effect of pipe friction losses.
Furthermore, the diameter of the pressure line considered in the previous sections was assumed 
to be constant.
Here, we will focus our attention on a more general situation, namely a pressure line
which is characterized by the coefficient functions $D$, $c$ and $f_\lambda$ as depicted in
Fig.~\ref{fig:figure_water_hammer_2_parameters}.
All the other parameters are chosen like in the previous water hammer problem
and also the boundary conditions and the time evolution of the control input $u$ 
remain the same.

\begin{figure}
\begin{center}
\includegraphics[scale=0.75]{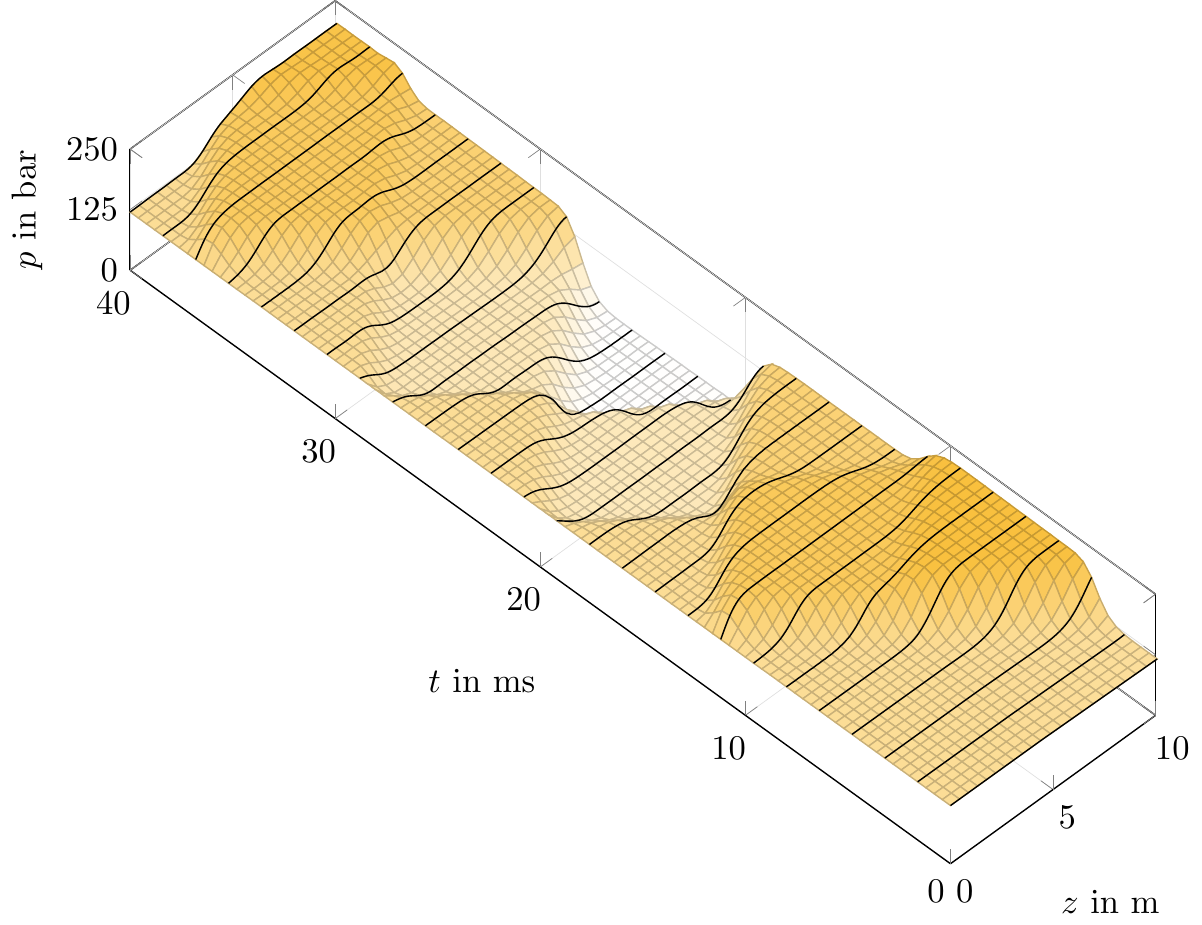}
\hspace{-3.5cm}
\includegraphics[scale=0.75]{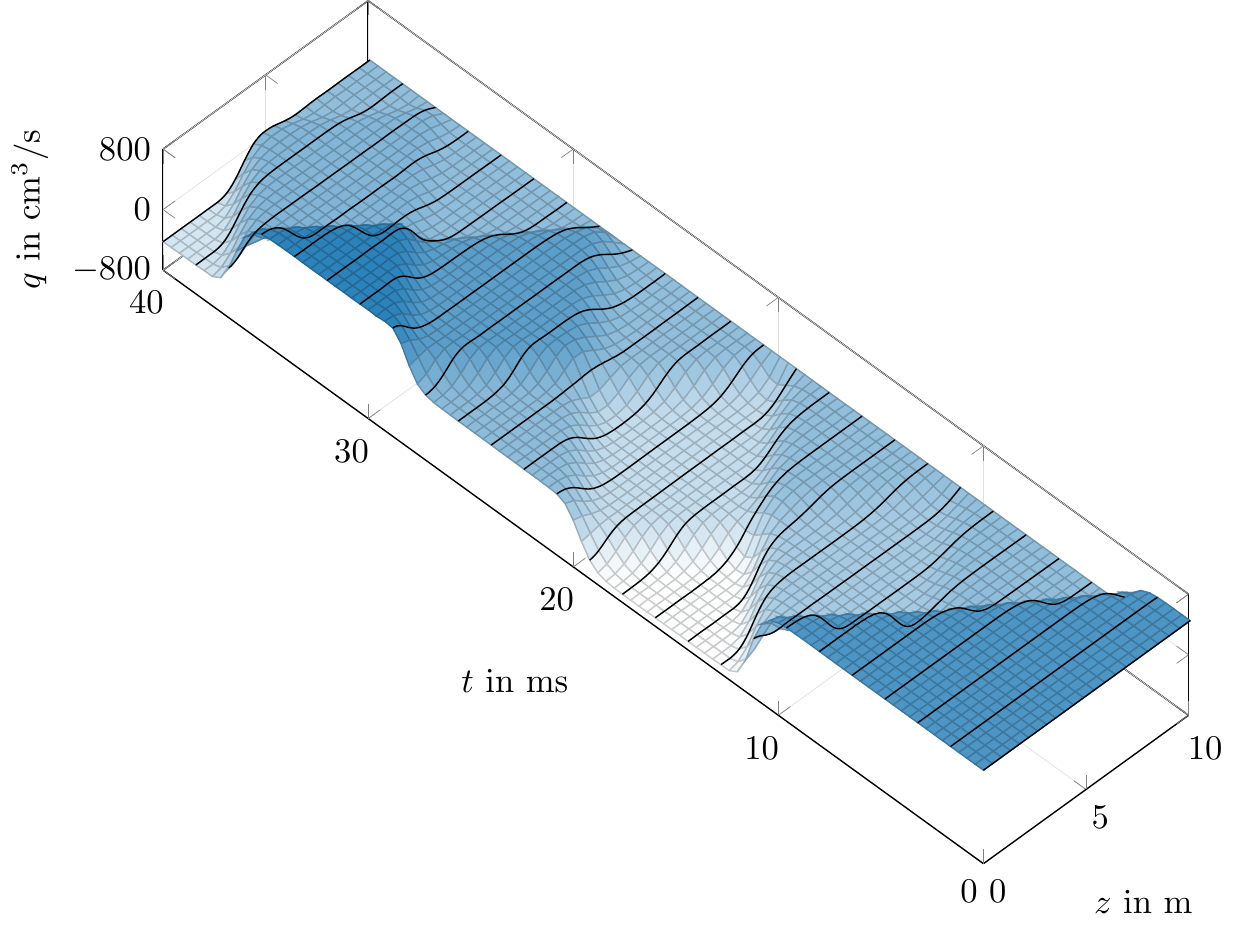}
\end{center}
\caption{
Time evolution of $p$ and $q$ corresponding to the second water hammer problem.
}
\label{fig:water_hammer_2_time_evolution}
\end{figure}

The pressure line consists of three segments (separated by dashed orange lines)
as needed to resolve the abrupt changes in the coefficient functions.
In the first numerical simulation, the segments are subdivided into $M_1=10$, $M_2=1$ and $M_3=10$
elements, see Fig.~\ref{fig:figure_water_hammer_2_parameters}.
Moreover, a polynomial degree of $N=3$ is assigned to all elements.
Please note that the element sizes vary slightly from one segment of the pressure line to another.
In order to propagate the numerical solution in time we utilize the classical Runge Kutta method
in combination with a time step size of $\triangle t = 0.2$\,ms.
The corresponding time evolution of $p$ and $q$ is shown in 
Fig.~\ref{fig:water_hammer_2_time_evolution} and exhibits much more complicated patterns 
compared to the first water hammer problem (cf. Fig.~\ref{fig:water_hammer_1_time_evolution}).

Taking a closer look at the derivation presented in Sec.~\ref{sec:numerical_method}, it is quite
obvious that it makes hardly any difference whether the SEM is used in combination with
constant or piecewise smooth coefficient functions 
as long as  abrupt changes in $D$, $c$, or $f_\lambda$
are taken into account by a properly adapted spatial mesh.
This is in strong contrast to the MOC which in case of a variable diameter $D$ 
performs significantly worse.
Also a non-zero pipe friction coefficient $f_\lambda$ deteriorates the accuracy of the MOC.
However, since pipe frictional losses are comparatively small, this effect becomes only visible
at very fine spatio-temporal discretizations.

For the reasons indicated above, it is basically impossible to obtain a sufficiently accurate
solution by means of the MOC which will then serve as a reference solution to investigate
the numerical convergence rates of the SEM.
Instead, we will show that for decreasing step sizes $(\triangle t)_\mathrm{MOC}$
the solutions of the MOC converge towards a reference solution provided by the SEM.
Here, it should be noted that
the spatial mesh sizes $(\triangle z)_\mathrm{MOC}^{(k)}$ are completely determined by the 
wave speeds on the different segments $k=1,2,3$ via the CFL condition \eqref{eq:CFL_MOC}.

\begin{figure}
\centering
\includegraphics[width=75mm]{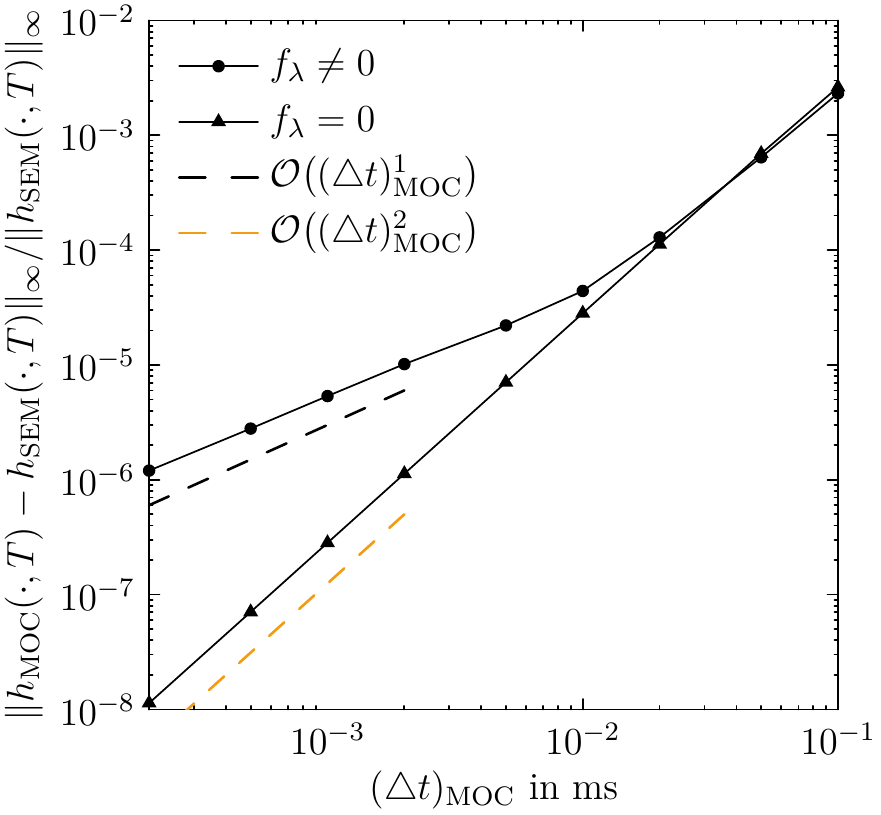}
\caption{
Relative differences between the numerical solutions of the MOC and the SEM corresponding to the second 	
water hammer problem evaluated at $T=20$\,ms.
For decreasing time step sizes $(\triangle t)_\mathrm{MOC}$, the solutions of the MOC converge towards a 
reference solution provided by the SEM.
The spatial mesh sizes $(\triangle z)_\mathrm{MOC}^{(k)}$ corresponding to the MOC are completely determined by 
$(\triangle t)_\mathrm{MOC}$ and the wave speeds $c_k$ on the different segments $k=1,2,3$ via the CFL 
condition \eqref{eq:CFL_MOC}.
Depending on whether pipe frictional losses are taken into account 
($f_\lambda \neq 0$) or neglected ($f_\lambda = 0$),
the relative differences decrease at a linear or quadratic rate, respectively.
}
\label{fig:convergence_moc}
\end{figure}

In order to compute a reference solution by means of the SEM,
the first, second and third segment is partitioned into $40$, $4$ and $40$ elements, respectively.
Moreover, a polynomial degree of $N_m=8$ is used for all elements $m=1,\dots,M$
and the time evolution is based on the \texttt{Matlab} function 
\texttt{ode45}~($\texttt{RelTol} = 2.5\times 10^{-14}$).
It is worth noting that a further refinement of the spatio-temporal discretization would not change the
relative differences shown in Fig.~\ref{fig:convergence_moc}.

It can clearly be seen that, for practically relevant time step sizes $(\triangle t)_\mathrm{MOC}$, 
the numerical solutions  corresponding to the MOC converge at a rate
of approximately $2$ towards the reference solution provided by the SEM.
However, for small time step sizes the convergence rate decreases to a value of approximately $1$.
This behavior agrees well with the numerical approximations underlying the MOC.
The derivation of the MOC is based on an integration of the characteristic equations 
along the positive and negative characteristics $dz/dt = c$ and $dz/dt = -c$, respectively.
To this end, variable cross-sectional areas are replaced by average values 
corresponding to two consecutive grid points, which causes the second order convergence rate
shown in Fig.~\ref{fig:convergence_moc}.
In order to integrate the nonlinear pipe frictional loss term, the expression
$q |q|$ is approximated by its value at the beginning of each time step.
While this simple approximation is necessary in order to make the MOC an explicit method,
it is also responsible for the first order convergence rate seen in 
Fig.~\ref{fig:convergence_moc}.
However, since pipe frictional losses are very small the effect is only visible at a low 
level of relative differences.
In fact, if pipe frictional losses are neglected ($f_\lambda = 0$) we obtain a uniform 
quadratic convergence rate, cf. Fig.~\ref{fig:convergence_moc}.

By means of the above simulations we have, at least implicitly, demonstrated that the
SEM works as expected also in the case of variable coefficient functions $D$, $c$ and $f_\lambda$.
Furthermore, we have investigated the numerical convergence behavior of the MOC.
To the best of our knowledge, and despite the fact that the MOC is the most widely used method
in the context of transient hydraulic simulations, 
the convergence analysis of the MOC presented above 
cannot be found in literature in this form.
We would like to emphasize that in the numerical example, 
the wave speeds and the lengths of the segments
were chosen in such a way that the CFL conditions in \eqref{eq:CFL_MOC} 
can be easily satisfied.
In a real-world example the lengths of the segments often 
have to be modified with respect to the original data, and hence, the numerical solutions 
of the MOC converge towards the solution of a (slightly) modified problem.

\section{Simulation of hydraulic transients in a pumped-storage power plant}
\label{sec:power_plant}

\begin{figure}
\centering
\includegraphics[scale=1.0]{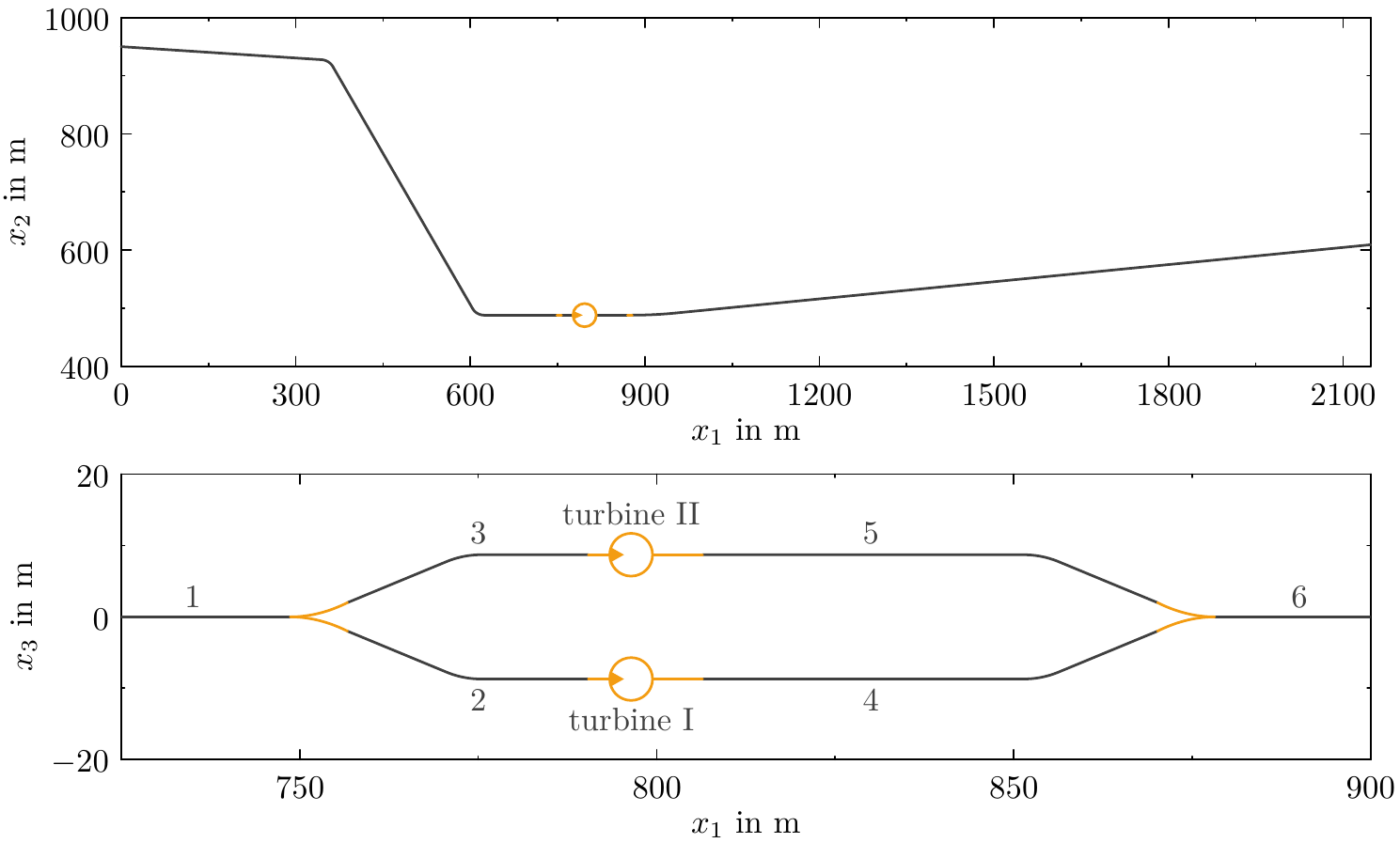}
\caption{
Spatial positions of the pressure lines, the bifurcations and the turbines
corresponding to the pumped-storage power plant considered in the numerical simulations of
Sec.~\ref{sec:power_plant}.
}
\label{fig:pipeline_system_locations}
\end{figure}

\begin{figure}
\centering
\includegraphics[scale=1.0]{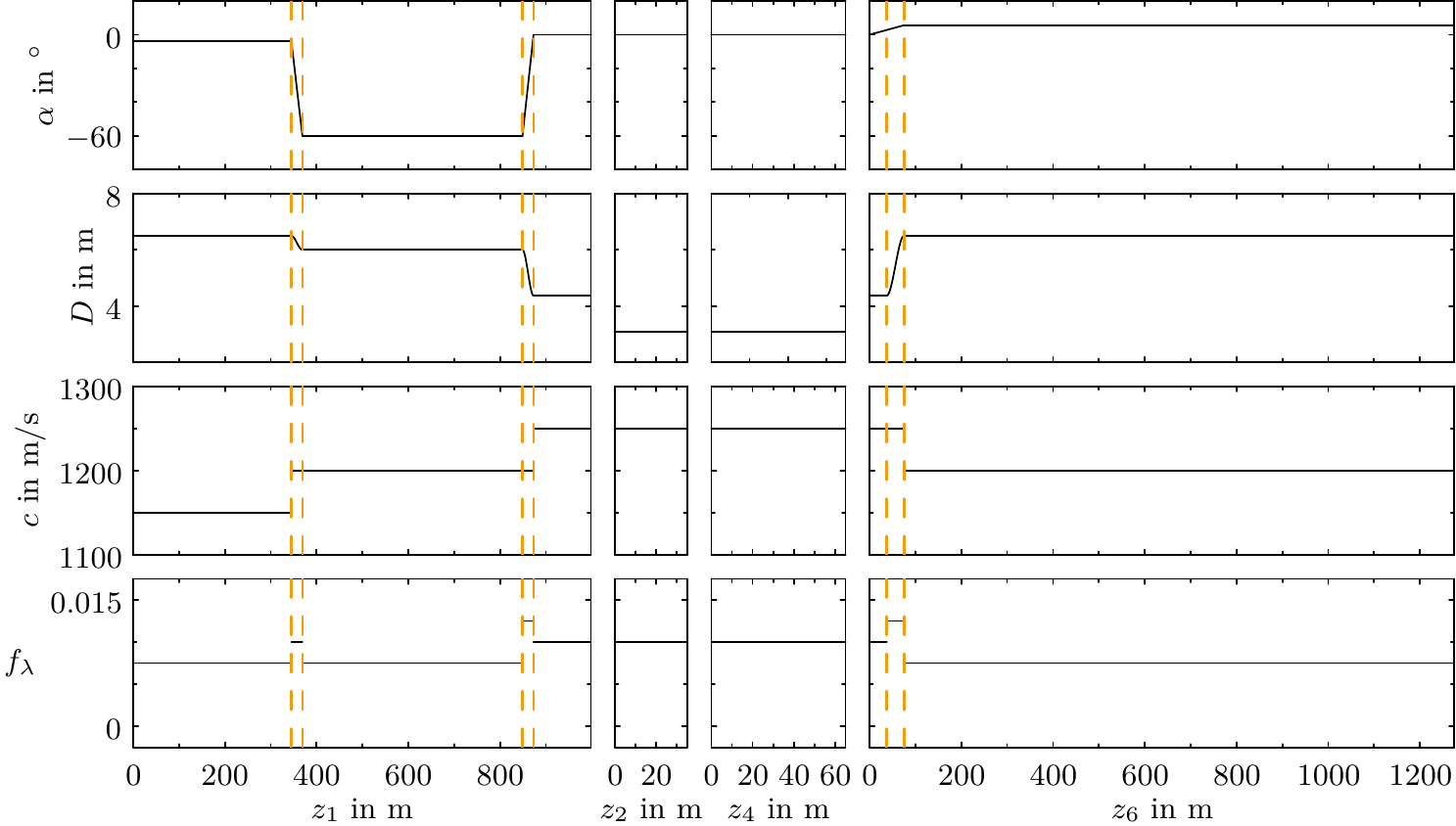}
\caption{
Parameter values characterizing the pipeline system of the pumped-storage power plant
depicted in Fig.~\ref{fig:pipeline_system_locations}.
Each of the six pressure lines is parameterized by its arclength $z_i \in [0,L_i]$, where $L_i$ denotes the length
of the $i$th pressure line, $i=1,\dots,6$.
The coefficient functions of the 3rd and 5th pressure line are assumed to
coincide with those of the 2nd and the 4th pressure line, respectively.
Different segments are separated by dashed orange lines.
In case of the SEM, each segment is discretized using a single element.
}
\label{fig:pipeline_system_parameters}
\end{figure}

Having demonstrated the accuracy and effectiveness of the SEM for the numerical simulation 
of wave propagation effects on a single pressure line,
we will now turn our attention to a coupled system of several pressure lines.
In particular, we want to simulate the transient behavior of pressure waves
in a large-scale pumped-storage power plant.
A detailed mathematical model of a variable speed pumped-storage power plant including 
a comprehensive description of the underlying hydraulic and electrical systems 
is presented in~\cite{schmidt_2017}.

The main reason for considering the pumped-storage power plant example is to study the applicability 
and the advantage of the SEM for a real-world control problem.
For the application of modern optimization-based control techniques like model predictive control, appropriate
numerical approximations which are computationally very efficient and at the same time
capable of accurately describing the time evolution of the pressure waves are a key element.

\subsection{Model description}

The spatial positions of the pressure lines, the bifurcations and the turbines are illustrated in 
Fig.~\ref{fig:pipeline_system_locations} using a $(x_1,x_2,x_3)$-coordinate frame.
Here, $x_2$ is oriented in opposite direction to the gravitational acceleration $g$.
Each of the pressure lines will be parameterized by its arclength $z_i \in [0,L_i]$,
with $L_i$ denoting the length of the $i$th pressure line, $i=1,\dots,6$.
Moreover, we assume that the 1st pressure line is connected to a reservoir at $z_1(0)$
and the 6th pressure line is connected to a reservoir
at $z_6(L_6)$.

Fig.~\ref{fig:pipeline_system_parameters} shows the coefficient functions 
$\alpha$, $D$, $c$ and $f_\lambda$
corresponding to the 1st, 2nd, 4th and 6th pressure line.
The coefficient functions of the 3rd and 5th pressure line
coincide with those of the 2nd and the 4th pressure line, respectively.
We note that the inclination $\alpha_i$ determines the height profile 
\[
x_{2,i}(z_i) = x_{2,i}^0 + \int_0^{z_i} \sin(\alpha_i(\tilde{z})) \,d\tilde{z}, \quad z_i \in [0,L_i]
\]
of the $i$th pressure line as a function of the local arclength $z_i$.
However, these height profiles are only needed for the conversion of pressure values to 
piezometric heads (and vice versa), see~\eqref{eq:relation_h_p}.
The time evolution of the piezometric head $h_i$ and the volume flow $q_i$ on the $i$th
pressure line is governed by~\eqref{eq:system_hq}, i.e.
\begin{equation}
\label{eq:system_hq_multiple}
\frac{\partial}{\partial t}
\begin{bmatrix}
h_{i}\\
q_{i}
\end{bmatrix}
+
\begin{bmatrix}
0 & c_i^2/(g A_i) \\
g A_i & 0
\end{bmatrix}
\frac{\partial}{\partial z}
\begin{bmatrix}
h_i\\
q_i
\end{bmatrix}
=
-
\begin{bmatrix}
0 \\
R_i q_i |q_i|
\end{bmatrix}
\end{equation}
for $i=1,\dots,6$.
In order to complete the description of the mathematical model, we
need to specify twelve boundary conditions.

Given the relatively short simulation periods considered below, it is reasonable to assume that
the filling levels of the reservoirs are virtually constant for all $t \in [0,T]$.
Therefore, the piezometric heads at the reservoirs
are fixed by Dirichlet boundary conditions
\begin{subequations}
\label{eq:boundary_conditions_reservoirs}
\begin{align}
h_1(0,t)   &= h_\mathrm{res, t}, \\
h_6(L_6,t) &= h_\mathrm{res, b},
\end{align}
\end{subequations}
where
\begin{align*}
h_\mathrm{res, t} &= p_\mathrm{res, t} / (\rho g) + x_{2,1}(0), \\
h_\mathrm{res, b} &= p_\mathrm{res, b} / (\rho g) + x_{2,6}(L_6)
\end{align*}
with $p_\mathrm{res, t}$ and $p_\mathrm{res, b}$ being the pressure
at the top and bottom reservoir, respectively. 

\begin{figure}
\centering
\includegraphics[width=85mm]{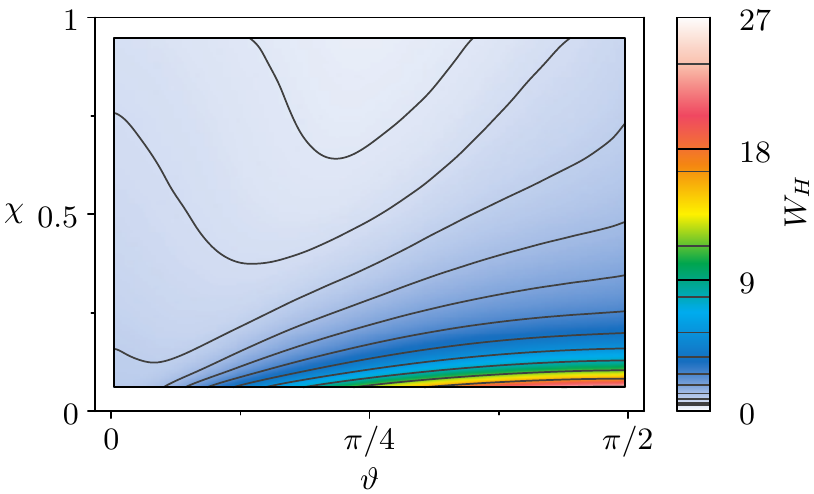}
\caption{Characteristic map $W_H$.}
\label{fig:wh}
\end{figure}

Furthermore, since the lengths of the bifurcations are short in comparison to the lengths 
of the pressure lines, it is fair to assume that the bifurcations can be well described by
the relations
\begin{equation}
\label{eq:boundary_conditions_bifurcations}
\begin{aligned}
q_1(L_1,t) &= q_2(0,t) + q_3(0,t), \\
h_2(0,t) &= h_1(L_1,t), \\
h_3(0,t) &= h_1(L_1,t), \\
h_4(L_4,t) &= h_6(0,t), \\
h_5(L_5,t) &= h_6(0,t), \\
q_6(0,t) &= q_4(L_4,t) + q_5(L_5,t).
\end{aligned}
\end{equation}
The remaining four boundary conditions are
\begin{equation}
\label{eq:boundary_conditions_turbines}
\begin{aligned}
q_2(L_2,t) &= q_\mathrm{I}, \\
q_3(L_3,t) &= q_\mathrm{II}, \\
q_4(0,t) &= q_\mathrm{I}, \\
q_5(0,t) &= q_\mathrm{II},
\end{aligned}
\end{equation}
where the volume flows through the turbines
$q_\mathrm{I}$ and $q_\mathrm{II}$
are given implicitly by the nonlinear algebraic equations \cite{schmidt_2017}
\begin{subequations}
\begin{align}
\begin{split}
\label{eq:turbine_1st}
0
&=
W_H(\chi_\mathrm{I}, \vartheta_\mathrm{I}) [q_\mathrm{I}/q_\mathrm{ref})^2 
+ (\omega_\mathrm{I}/\omega_\mathrm{ref})^2] h_\mathrm{ref}
-\frac{q_\mathrm{I}^2}{2 g} [(1/A_2(L_2))^2 - (1/A_4(0))^2] \\
&\qquad
-(h_2(L_2,t) - h_4(0,t)),
\end{split}
\\
\begin{split}
\label{eq:turbine_2nd}
0
&=
W_H(\chi_\mathrm{II}, \vartheta_\mathrm{II}) [(q_\mathrm{II}/q_\mathrm{ref})^2 
+ (\omega_\mathrm{II}/\omega_\mathrm{ref})^2] h_\mathrm{ref}
-\frac{q_\mathrm{II}^2}{2 g} [(1/A_3(L_3))^2 - (1/A_5(0))^2] \\
&\qquad
-(h_3(L_3,t) - h_5(0,t)).
\end{split}
\end{align}
\end{subequations}
Here, $\omega_\mathrm{I}$ and $\omega_\mathrm{II}$ denote angular velocities corresponding to the two 
turbines and
\[
\vartheta_\mathrm{I} = 
\arctan \Big( \frac{q_\mathrm{I} / q_\mathrm{ref}} { \omega_\mathrm{I} / \omega_\mathrm{ref} } \Big),
\quad
\vartheta_\mathrm{II} = 
\arctan \Big( \frac{q_\mathrm{II} / q_\mathrm{ref}} { \omega_\mathrm{II} / \omega_\mathrm{ref} } \Big)
\]
are dimensionless variables which, together with the guide vane positions $\chi_\mathrm{I}$
and $\chi_\mathrm{II}$, are used to evaluate the characteristic map $W_H$ 
shown in Fig.~\ref{fig:wh}.
We note that the above equations describe two Francis turbines which are assumed to be of 
the same type and, therefore, identical parameters
$h_\mathrm{ref}$, $q_\mathrm{ref}$ and $\omega_\mathrm{ref}$ are used both in~\eqref{eq:turbine_1st}
and~\eqref{eq:turbine_2nd}.
Moreover, in order to keep the model as simple as possible, we assume that the angular velocities
of the turbines are held constant 
at $\omega_\mathrm{I} = \omega_\mathrm{II} = \omega_\mathrm{ref}$ by means of a subordinate 
controller.

\subsection{Semi-discretization}

According to~\eqref{eq:semi_discr_hq}, the semi-discretizations of the
systems given in~\eqref{eq:system_hq_multiple} read
\begin{subequations}
\label{eq:semi_discr_plant}
\begin{equation}
\label{eq:semi_discr_plant_odes}
\begin{aligned}
M_{\epsilon, i} \frac{d \bm{h}_i}{dt}
&=
S_i \bm{q}_i + q_{i,1}^* \bm{e}_{i,1} - q_{i,J_i}^* \bm{e}_{i,J_i}, \\
M_{\mu,i} \frac{d \bm{q}_i}{dt}
&=
S_i \bm{h}_i + h_{i,1}^* \bm{e}_{i,1} - h_{i,J_i}^* \bm{e}_{i,J_i} - M_{r,i} \bm{q}_i |\bm{q}_i|
\end{aligned}
\end{equation}
for $i=1,\dots,6$.
The boundary conditions in \eqref{eq:boundary_conditions_reservoirs}, \eqref{eq:boundary_conditions_bifurcations} and \eqref{eq:boundary_conditions_turbines}
are realized by means of the numerical 
flux components (cf. Tab.~\ref{tab:dirichlet_boundary_conditions})
\begin{equation}
\label{eq:semi_discr_plant_flux_values}
\begin{aligned}
q_{1,1}^* &= q_{1,1} - g A_1(0) ( h_{1,1} - h_\mathrm{res,t} ) / c_1(0), 
&
h_{1,1}^* &= h_\mathrm{res,t},
\\
q_{6,J_6}^* &= q_{6,J_6} - g A_6(L_6) (h_\mathrm{res,b} - h_{6,J_6}) / c_6(L_6),
&
h_{6,J_6}^* &= h_\mathrm{res,b}, 
\\
q_{1,J_1}^* &= q_{2,1} + q_{3,1},
&
h_{1,J_1}^* &= h_{1,J_1} - c_{1}(L_1) ((q_{2,1} + q_{3,1})
\\
& & &\qquad - q_{1,J_1}) / (g A_1(L_1)), \\
q_{2,1}^* &= q_{2,1} + g A_2(0) (h_{1,J_1} - h_{2,1}) / c_2(0),
&
h_{2,1}^* &= h_{1,J_1}, 
\\
q_{3,1}^* &= q_{3,1} + g A_3(0) (h_{1,J_1} - h_{3,1}) / c_3(0),
&
h_{3,1}^* &= h_{1,J_1},
\\
q_{4,J_4}^* &= q_{4,J_4} - g A_4(L_4) ( h_{6,1} - h_{4,J_4} ) / c_4(L_4),
&
h_{4,J_4}^* &= h_{6,1}, 
\\
q_{5,J_5}^* &= q_{5,J_5} - g A_5(L_5) ( h_{6,1} - h_{5,J_5} ) / c_5(L_5),
&
h_{5,J_5}^* &= h_{6,1}, 
\\
q_{6,1}^* &= q_{4,J_4} + q_{5,J_5},
&
h_{6,1}^* &= h_{6,1} + c_{6}(0) ( (q_{4,J_4} + q_{5,J_5})
\\
& & &\qquad - q_{6,1}) / (g A_{6}(0)),
\\
q_{2,J_2}^* &= q_\mathrm{I},
&
h_{2,J_2}^* &= h_{2,J_2} - c_{2}(L_2) ( q_\mathrm{I} - q_{2,J_2} ) / (g A_{2}(L_2)),
\\
q_{4,1}^* &= q_\mathrm{I},
&
h_{4,1}^* &= h_{4,1} + c_4(0) ( q_\mathrm{I} - q_{1,4} ) / (g A_4(0)),
\\
q_{3,J_3}^* &= q_\mathrm{II},
&
h_{3,J}^* &= h_{3,J_3} - c_3(L_3) ( q_\mathrm{II} - q_{3,J_3} ) / (g A_3(L_3)),
\\
q_{5,1}^* &= q_\mathrm{II},
&
h_{5,1}^* &= h_{5,1} + c_5(0) ( q_\mathrm{II} - q_{5,1} ) / (g A_5(0)),
\end{aligned}
\end{equation}
where $q_\mathrm{I}$ and $q_\mathrm{II}$ are given implicitly by
\begin{equation}
\begin{aligned}
\label{eq:semi_discr_plant_algebraic_eq}
0
&=
W_H(\chi_\mathrm{I}, \vartheta_\mathrm{I}) [(q_\mathrm{I}/q_\mathrm{ref})^2 + 1] h_\mathrm{ref}
-q_\mathrm{I}^2 [(1/A_2(L_2))^2 - (1/A_4(0))^2] / (2 g)
-(h_{2,J_2}^* - h_{4,1}^*),
\\
0
&=
W_H(\chi_\mathrm{II}, \vartheta_\mathrm{II}) [(q_\mathrm{II}/q_\mathrm{ref})^2 + 1] h_\mathrm{ref}
-q_\mathrm{II}^2 [(1/A_3(L_3))^2 - (1/A_5(0))^2] / (2 g)
-(h_{3,J_3}^* - h_{5,1}^*).
\end{aligned}
\end{equation}
\end{subequations}
Here, it is important to note that $h_2(L_2,t), h_4(0,t)$ in~\eqref{eq:turbine_1st}
and $h_3(L_3,t), h_5(0,t)$ in~\eqref{eq:turbine_2nd}
have been replaced by $h_{2,J_2}^*$, $h_{4,1}^*$ and $h_{3,J_3}^*$, $h_{5,1}^*$, respectively.
Another possibility would be to use the nodal values $h_{2,J_2}$, $h_{4,1}$ and 
$h_{3,J_3}$, $h_{5,1}$ directly.
However, as in the case of the nonlinear valve boundary condition considered in 
Sec.~\ref{sec:water_hammer_simulation_1},
the latter choice requires a much smaller time step size $\triangle t$ in order to ensure numerical stability in combination with explicit time stepping methods.

\subsection{Numerical simulation}
\label{sec:numerical_simulation_power_plant}

We now consider a numerical simulation of the pumped-storage power plant introduced above.
In particular, we want to demonstrate that the SEM is able to reproduce a solution 
provided by the MOC using a minimum number of spatial grid points.
In this context, it is important to note that the relative errors
reported in Sec.~\ref{sec:convergence_analysis} 
are disproportionally small when compared to the expected errors of the underlying model.
Our long-term objective is to operate pumped-storage power plants 
using model predictive control (MPC).
In such applications, it is crucial to have a low-dimensional system 
representation available, which describes the dynamic behavior sufficiently well while remaining numerically efficient enough to be evaluated in real time.

The lengths of the pressure lines shown in Fig.~\ref{fig:pipeline_system_parameters} represent
realistic values but have been modified slightly in order to be able to satisfy the 
CFL conditions \eqref{eq:CFL_MOC} required by the MOC.
The largest time step size which satisfies all CFL conditions simultaneously 
is $(\triangle t)_\mathrm{MOC} = 2\times 10^{-3}$\,s.
Since the spatial mesh sizes of the MOC are determined by~\eqref{eq:CFL_MOC}
it follows that $J_\mathrm{MOC} = 2 \times 1036$ spatial grid points are needed for the discretization of 
the physical quantities $h$ and $q$.
Note that an even finer spatio-temporal discretization would not change the
numerical differences reported below.

In case of the SEM, we employ a single element per segment, see Fig.~\ref{fig:pipeline_system_parameters}.
Thus, in the given example, segments and elements coincide.
While our implementation allows for arbitrary subdivisions of the segments, it turned out
that single elements of comparatively high order are more efficient when it comes to reaching medium level 
accuracies using a minimum number of spatial grid points.
The polynomial degrees of the element basis functions corresponding to the different 
pressure lines and segments are
$\mathcal{N}_1 = (3,1,3,1,2)$,
$\mathcal{N}_2 = \mathcal{N}_3 = 1$,
$\mathcal{N}_4 = \mathcal{N}_5 = 2$
and
$\mathcal{N}_6 = (1,1,8)$.
Thus, $J_\mathrm{SEM} = 2 \times 32$ degrees of freedom are used to represent the approximations 
of $h$ and $q$, cf.~\eqref{eq:nr_of_basis_functions}.
Since \eqref{eq:semi_discr_plant} constitutes a semi-explicit system of 
differential algebraic equations (DAE) of index 1, there are basically two
established approaches to realize the time integration.
The first approach consists of solving the algebraic equations 
\eqref{eq:semi_discr_plant_algebraic_eq} locally for the algebraic 
variables $q_\mathrm{I}$ and $q_\mathrm{II}$.
Hence, the remaining system of equations 
\eqref{eq:semi_discr_plant_odes} and \eqref{eq:semi_discr_plant_flux_values} can be treated using explicit time stepping methods like the classical Runge Kutta method or \texttt{ode45} 
as implemented in the \texttt{Matlab} scripting language.
Alternatively, the whole system is treated as a single DAE system which can be integrated in time using implicit methods like \texttt{ode15s} which is also an integral part of the \texttt{Matlab} programming environment.

\begin{figure}
\centering
\includegraphics[scale=1]{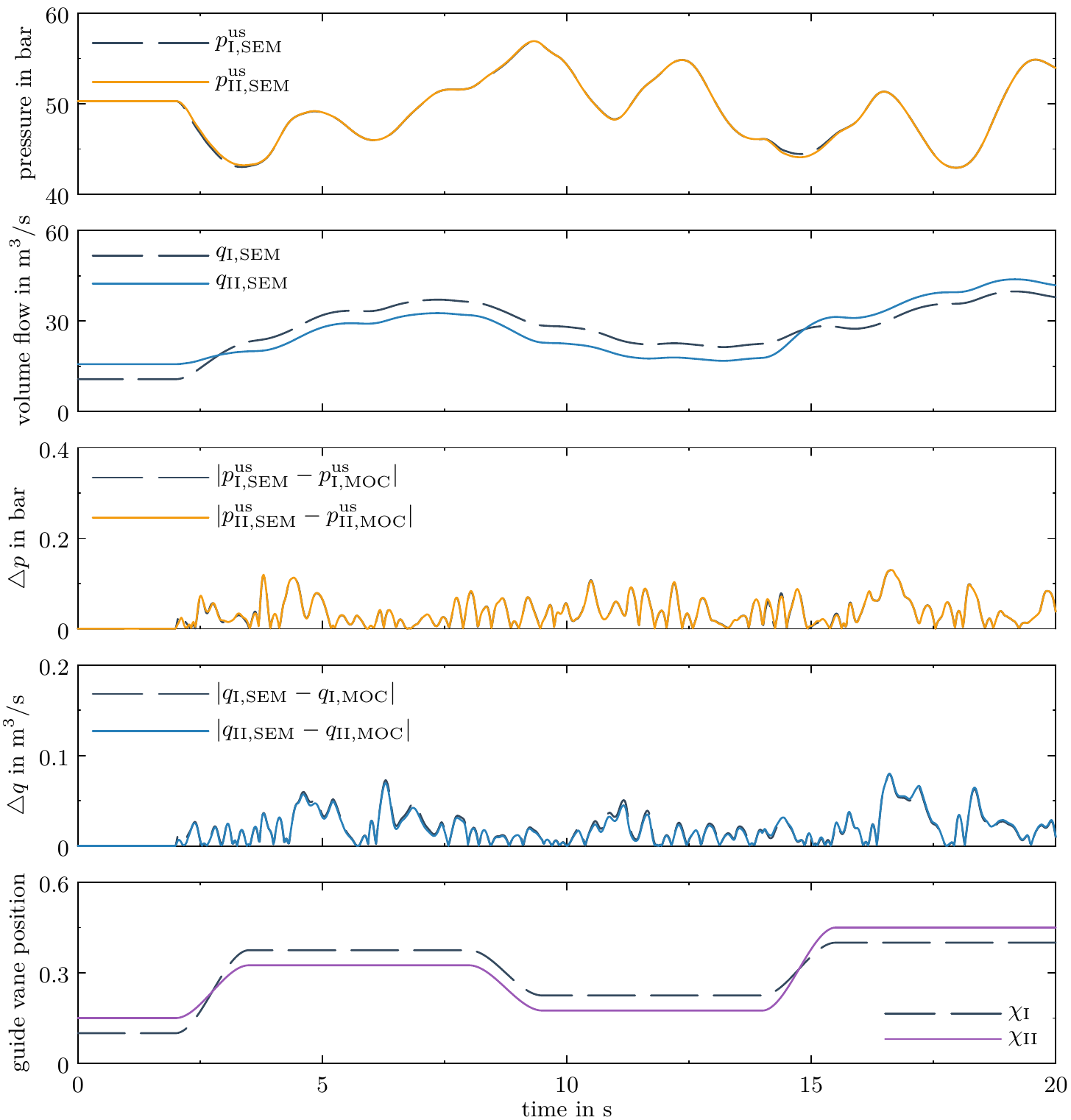}
\caption{
Numerical simulation of the pumped-storage power plant considered in Sec.~\ref{sec:numerical_simulation_power_plant}.
The guide vane positions $\chi_\mathrm{I}$ and $\chi_\mathrm{II}$ are varied as
illustrated in the last row.
The time evolution of the pressure at the upstream side of the first and second turbine is shown 
in the first row and the time evolution of the volume flows through the turbines is depicted in the second row.
As can be seen from the third and fourth row, the (absolute) differences between the solutions 
of the SEM and the MOC are well below $5 \times 10^{-3}$ for all times $t \in [0,T]$.
}
\label{fig:signatures_system_part_1}
\end{figure}

The numerical simulation is based on the following set of physical parameters:
$p_\mathrm{res,t}    = 5\,\mathrm{bar}$, 
$p_\mathrm{res,b}    = 2\,\mathrm{bar}$, 
$h_\mathrm{ref}      = 313.4\,\mathrm{m}$,
$q_\mathrm{ref}      = 44.7\,\mathrm{m}^3/\mathrm{s}$,
and
$\omega_\mathrm{ref} = 42.1\,\mathrm{rad}/\mathrm{s}$.
Moreover, we assume that the initial state of the system is given by a stationary solution of
\eqref{eq:semi_discr_plant} corresponding to the initial guide vane positions
$\chi_\mathrm{I}(t=0)$ 
and $\chi_\mathrm{II}(t=0)$.
The guide vane positions $\chi_\mathrm{I}$ and $\chi_\mathrm{II}$ will be varied as
shown in the last row of Fig.~\ref{fig:signatures_system_part_1} for the simulation 
period $[0,T]$ with $T=20$\,s.

Due to the restrictions imposed by the CFL conditions, the MOC requires $10\,000$ time steps
in order to compute the complete time evolution.
The elapsed calculation time amounts to approximately $4.4$ seconds.
In contrast, if the time evolution is computed by means of the SEM and 
\texttt{ode45}~($\texttt{RelTol} = 2.5\times 10^{-6}$)
the required calculation time amounts to approximately $1.1$ seconds.
Using \texttt{ode15s}~($\texttt{RelTol} = 2.5\times 10^{-6}$), the computing time can be 
even further reduced to roughly $0.75$ seconds.

The first row of Fig.~\ref{fig:signatures_system_part_1} illustrates the time evolution 
of the pressure at the upstream side of the first and second turbine, respectively.
Due to the short lengths of the 2nd, 3rd, 4th and 5th pressure line the differences of both graphs are 
comparatively small.
The second row shows the time evolution of the 
volume flows through the turbines.
It can be clearly seen that even within time intervals where the guide vane positions 
are held constant the volume flows in the turbines are far from being constant as well.
In fact, it would take much longer time intervals ($\approx 20$\,s)
until the system becomes (approximately) stationary again.
Additionally, the third and fourth row of Fig.~\ref{fig:signatures_system_part_1} demonstrate that
the corresponding (absolute) differences between the solutions of the SEM and the MOC
are well below $5 \times 10^{-3}$ for all times $t \in [0,T]$. 

\begin{figure}
\centering
\includegraphics[scale=1]{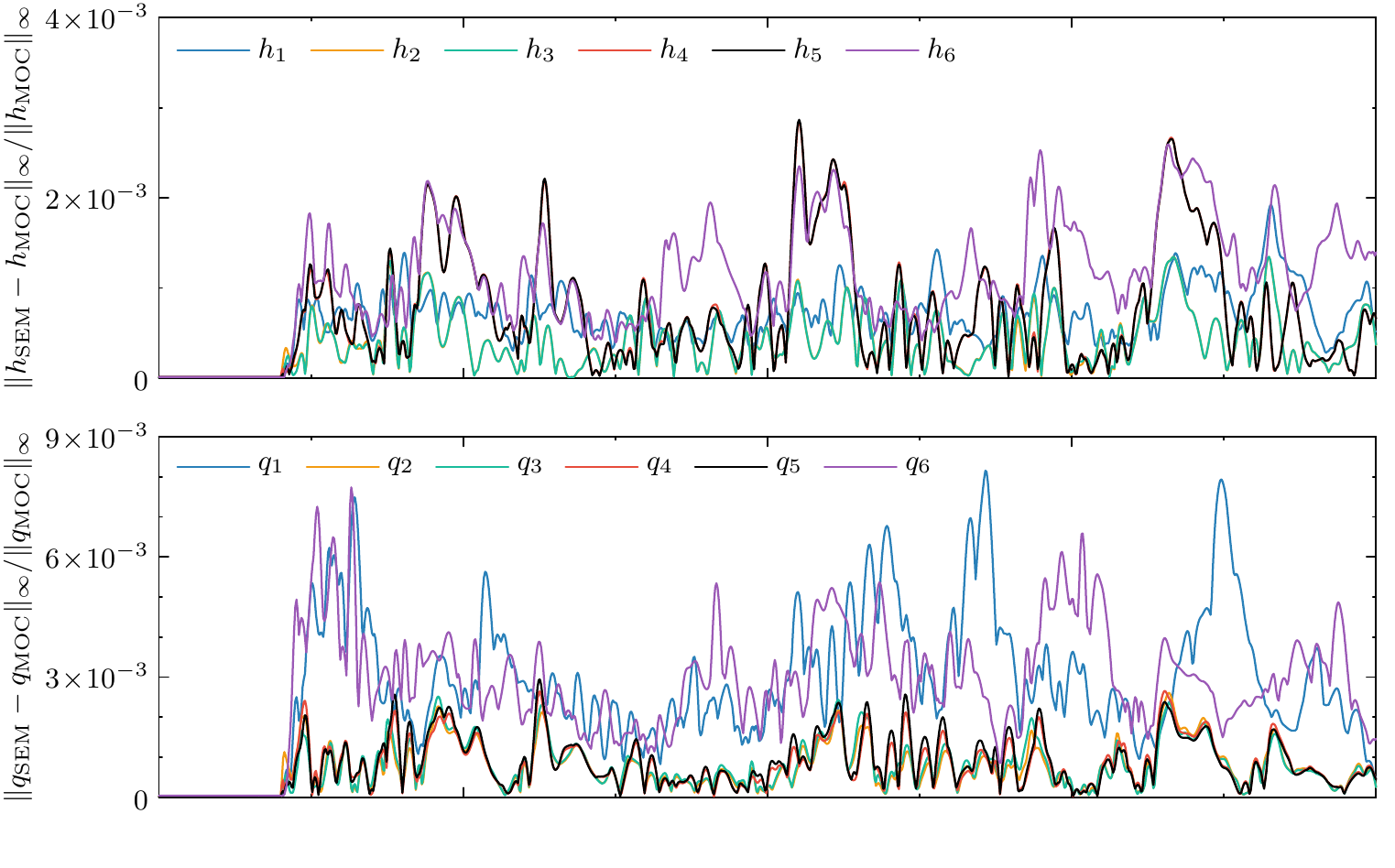}
\caption{
Relative differences between the solutions corresponding to the SEM and the MOC for the numerical 
simulation of the pumped-storage power plant considered in Sec.~\ref{sec:numerical_simulation_power_plant}.
The relative differences of the piezometric heads and the volume flows 
are shown as a function of time and separately for all six pressure lines.
}
\label{fig:signatures_system_part_2}
\end{figure}

So far, we have only considered the differences of the numerical solutions
corresponding to the SEM and the MOC at or in the two Francis turbines.
However, in many applications the numerical solutions are required to be accurate along the whole pipeline system.
For this reason, the relative differences between the numerical solutions
corresponding to the SEM and the MOC are given in Fig.~\ref{fig:signatures_system_part_2}.
More precisely, the relative differences of the piezometric heads and the volume flows 
are shown as a function of time and separately for all six pressure lines.
Quite obviously,
the relative differences of the piezometric heads $h_1, \dots, h_6$ and the volume 
flows $q_1, \dots, q_6$ are well below $1 \times 10^{-2}$ for all times $t \in [0,T]$
and hence the given level of accuracy is significantly smaller than the expected overall modeling error.
We note that, in the present case, the time evolution corresponding to the SEM was calculated using 
the \texttt{Matlab} function \texttt{ode45}.
However, the second approach based on \texttt{ode15s} yields practically the same results.

\begin{figure}
\begin{center}
\includegraphics[scale=0.75]{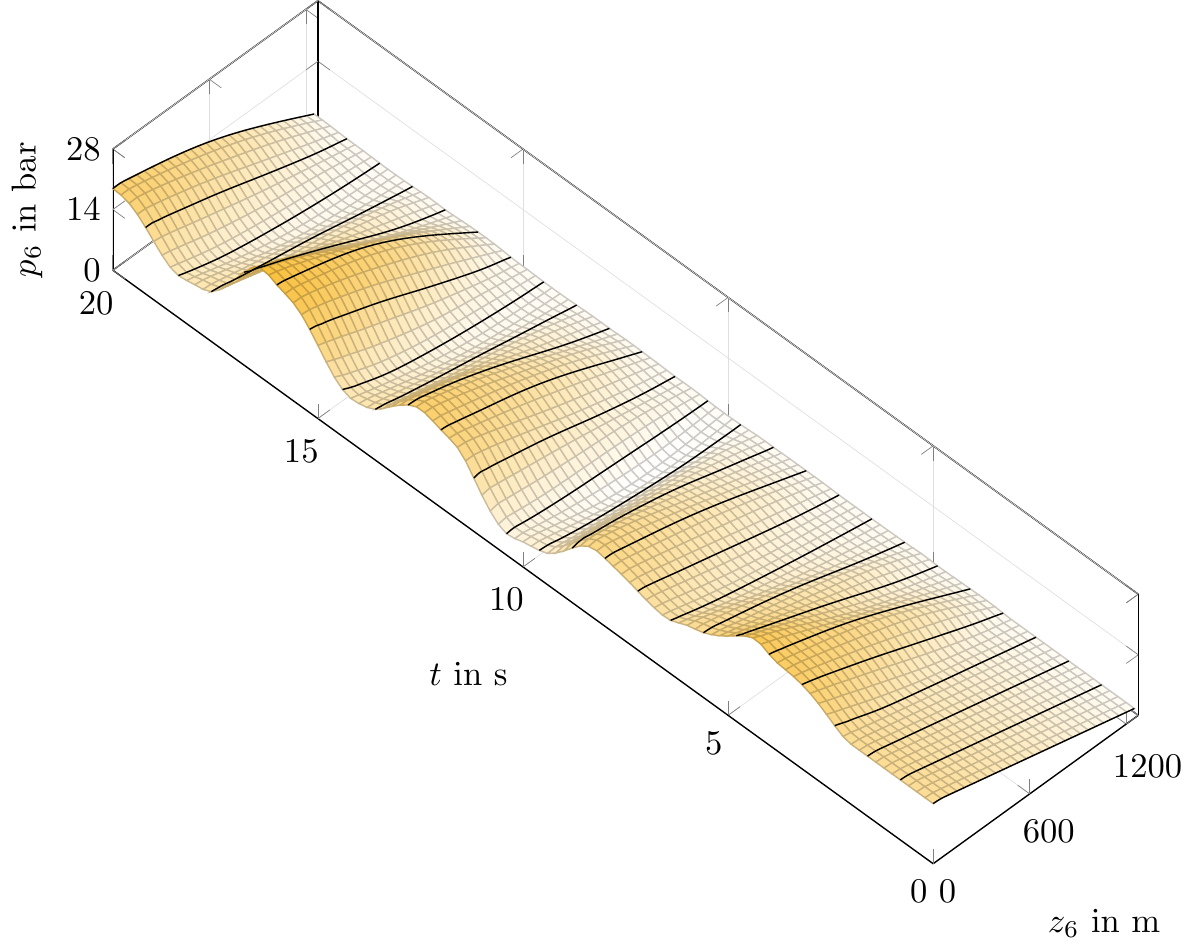}
\hspace{-3.5cm}
\includegraphics[scale=0.75]{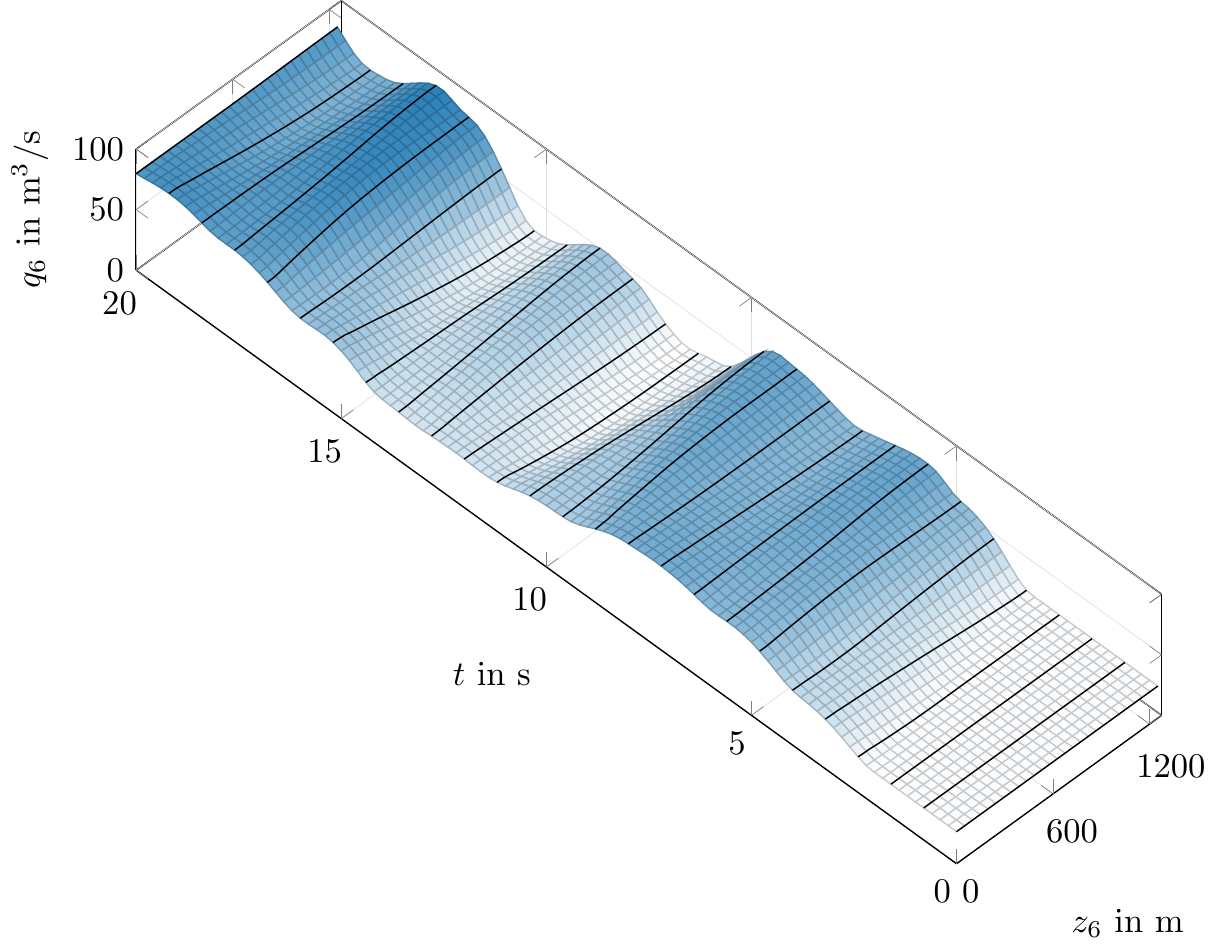}
\end{center}
\caption{
Time evolution of $p_6$ and $q_6$ corresponding to a numerical simulation of the pumped-storage power plant 
depicted in Fig.~\ref{fig:pipeline_system_locations}.
}
\label{fig:time_evolution_p_q_6}
\end{figure}

\begin{figure}[htb]
\centering
\includegraphics[scale=1]{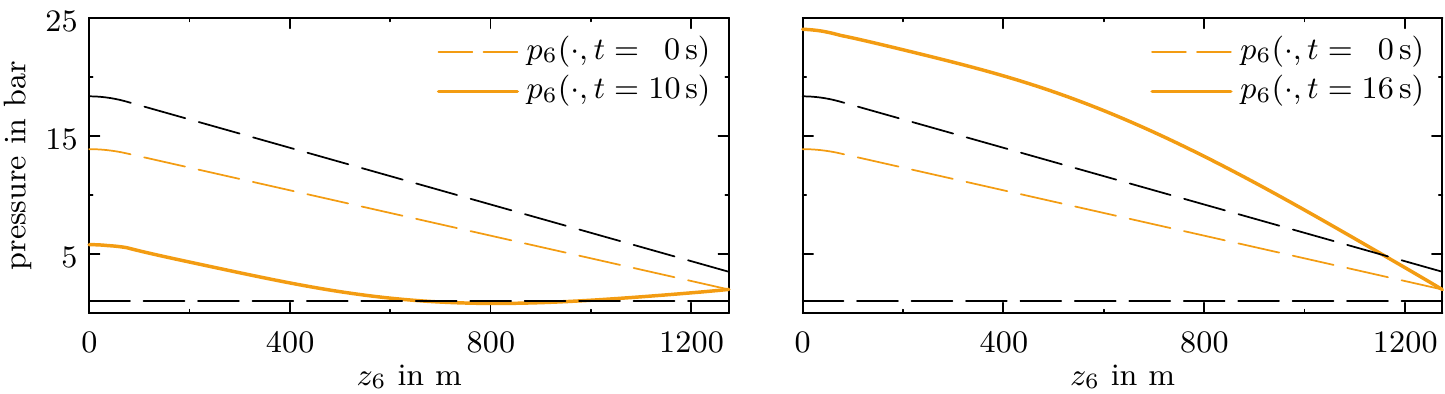}
\caption{
Snapshots of the pressure distribution along the 6th pressure line of the pumped-storage 
power plant considered in the numerical simulation of Sec.~\ref{sec:power_plant}.
}
\label{fig:snapshots_system_scenario_1}
\end{figure}

Of course, one could object that faster variations of the guide vane positions would cause stronger pressure fluctuations which could not be resolved by the low-dimensional discretization of the SEM outlined above.
However, it turns out that the variations of $\chi_\mathrm{I}$ and $\chi_\mathrm{II}$
already represent a possible worst case scenario.
To make this clear, we consider the time evolution of the pressure on the 6th pressure line
as illustrated in Fig.~\ref{fig:time_evolution_p_q_6}.
Due to its length the 6th pressure line shows the most pronounced variations of the pressure.
Moreover, due to its spatial position, the pressure on the 6th pressure line is most likely to
approach zero, which, with respect to cavitation, is a critical scenario.
In fact, we find that the time evolution of $p$ and $q$ is comparatively smooth.
However, Fig.~\ref{fig:snapshots_system_scenario_1} unveils that at $t=10$\,s
the pressure becomes very close to zero (locally).
In a real system, the pressure should never fall below $1$\,bar (lower dashed line) and due 
to model inaccuracies even larger lower bounds are realistic.
At the same time, the pressure along the pressure lines is required to remain below a reasonable 
upper bound, which is defined in terms of the pressure distribution of the 
stationary state (orange dashed line).
However, Fig.~\ref{fig:snapshots_system_scenario_1} shows that  even a generously dimensioned 
upper bound (upper dashed line) is dramatically exceeded at $t=16$\,s.

In optimal or model predictive control applications, these lower and upper bounds for 
the pressure distributions must be systematically taken into account.
In this context, the most rigorous approach is based on
direct transcription methods \cite{betts_2010, gerdts_2012}. 
However, in order to keep the numerical effort as low as possible,
direct transcription methods are dependent on a low dimensional system representation.
Comparing the system size of the MOC ($J_\mathrm{MOC} = 2 \times 1036$) with the system 
size of the SEM ($J_\mathrm{SEM} = 2 \times 32$),
it is obvious that the SEM will give a 
significant advantage in this field of applications.

\section{Conclusions and outlook}
\label{sec:conclusion}

In summary, this paper shows that the spectral element method (SEM)
represents an efficient and flexible method for transient simulations of hydraulic systems,
in particular also for the use in optimization-based control applications.
By means of the SEM it is possible to compute highly accurate numerical solutions 
of the water hammer problem using modest computational resources.
As a high order method, the SEM requires a minimum number of dynamical states
in order to yield excellent approximations of
sufficiently smooth pressure and volume flow fluctuations.
In fact, we demonstrated that the hydraulic system of a large scale pumped-storage power plant
can be well approximated using a low-dimensional system representation 
which consists of only a few dozen dynamical states.
The corresponding semi-discretization can be integrated in time using standard 
time stepping strategies and is therefore well suited for optimal and model predictive control applications.
In particular, model predictive control of pumped-storage power plants,
while respecting lower and upper pressure bounds, is within reach and will be presented in
a subsequent publication.

\section*{Acknowledgements}
The authors would like to thank M. Egretzberger and M. Meusburger from Andritz Hydro GmbH for 
the fruitful cooperation. 
This work was supported by the Austrian Research Promotion Agency (FFG), Grant No.: 849267. 
Furthermore, the financial support by the Austrian Federal Ministry of Science, Research and 
Economy and the National Foundation for Research, Technology and Development is 
gratefully acknowledged.

\bibliography{pipeline_transients}

\end{document}